\documentclass[10pt,journal,compsoc]{IEEEtran}

\makeatletter
\def\ps@IEEEtitlepagestyle{
  \def\@oddfoot{\mycopyrightnotice}
  \def\@evenfoot{}
}

\def\mycopyrightnotice{
  {\footnotesize
  \begin{minipage}{\textwidth}
  \centering
  Copyright~\copyright~2020 IEEE.  Personal use of this material is permitted.  Permission from IEEE must be obtained for all other uses, in any current or future media, including reprinting/republishing this material for advertising or promotional purposes, creating new collective works, for resale or redistribution to servers or lists, or reuse of any copyrighted component of this work in other works.
  \end{minipage}
  }
}

\ifCLASSOPTIONcompsoc
  \usepackage[nocompress]{cite}
\else
  \usepackage{cite}
\fi

\ifCLASSINFOpdf
   \usepackage[pdftex]{graphicx}
   \graphicspath{ {./images/} }
    \usepackage[outdir=./]{epstopdf}
\else
  \usepackage[dvips]{graphicx}
\fi

\usepackage{booktabs}

\usepackage{amsmath}

\usepackage{algorithm}
\usepackage[noend]{algpseudocode}

\usepackage{amssymb}
\usepackage{multirow}
\usepackage{enumitem}

\usepackage[utf8]{inputenc}
\usepackage[english]{babel}
\usepackage{amsthm}
\theoremstyle{definition}
\newtheorem{definition}{Definition}
\usepackage{xcolor}
\usepackage{enumitem}

\usepackage[normalem]{ulem}

\usepackage{array}

\usepackage{url}

\ifCLASSOPTIONcompsoc
	\usepackage[caption=false,font=footnotesize,labelfont=sf,textfont=sf]{subfig}
\else
	\usepackage[caption=false,font=footnotesize]{subfig}
\fi

\begin{document}
%
% paper title
% Titles are generally capitalized except for words such as a, an, and, as,
% at, but, by, for, in, nor, of, on, or, the, to and up, which are usually
% not capitalized unless they are the first or last word of the title.
% Linebreaks \\ can be used within to get better formatting as desired.
% Do not put math or special symbols in the title.
%\title{Long-term Performance Discovery for IaaS Provider Selection}
\title{Long-term IaaS Selection using Performance Discovery}
%
%
% author names and IEEE memberships
% note positions of commas and nonbreaking spaces ( ~ ) LaTeX will not break
% a structure at a ~ so this keeps an author's name from being broken across
% two lines.
% use \thanks{} to gain access to the first footnote area
% a separate \thanks must be used for each paragraph as LaTeX2e's \thanks
% was not built to handle multiple paragraphs
%
%
%\IEEEcompsocitemizethanks is a special \thanks that produces the bulleted
% lists the Computer Society journals use for "first footnote" author
% affiliations. Use \IEEEcompsocthanksitem which works much like \item
% for each affiliation group. When not in compsoc mode,
% \IEEEcompsocitemizethanks becomes like \thanks and
% \IEEEcompsocthanksitem becomes a line break with idention. This
% facilitates dual compilation, although admittedly the differences in the
% desired content of \author between the different types of papers makes a
% one-size-fits-all approach a daunting prospect. For instance, compsoc 
% journal papers have the author affiliations above the "Manuscript
% received ..."  text while in non-compsoc journals this is reversed. Sigh.

\author{Sheik Mohammad Mostakim Fattah,
        Athman Bouguettaya, \IEEEmembership{Fellow,~IEEE}, 
    and Sajib Mistry
\IEEEcompsocitemizethanks{\IEEEcompsocthanksitem S. Fattah and A. Bouguettaya are with the School of Computer Science, University of Sydney, Australia.
% note need leading \protect in front of \\ to get a newline within \thanks as
% \\ is fragile and will error, could use \hfil\break instead.
\protect \\ E-mail: \{sheik.fattah, athman.bouguettaya, sajib.mistry\}@sydney.edu.au
\newline
S.~Mistry is with the School of Electrical Eng, Computer and Math Science, Curtin University, Australia. Email: sajib.mistry@curtin.edu.au.
% \IEEEcompsocthanksitem H. Dong is with the School of Science, RMIT University, Australia. E-mail: hai.dong@rmit.edu.au
% \IEEEcompsocthanksitem A. Erradi is with the Department of Computer Science and Engineering, Qatar University, Qatar. E-mail: erradi@qu.edu.qa
% note need leading \protect in front of \\ to get a newline within \thanks as
% \\ is fragile and will error, could use \hfil\break instead.

}% <-this % stops a space
\thanks{}}

\IEEEtitleabstractindextext{%
\begin{abstract}
We propose a novel framework to select IaaS providers according to a consumer's long-term performance requirements. The proposed framework leverages free short-term trials to discover the unknown QoS performance of IaaS providers. We design a temporal skyline-based filtering method to select candidate IaaS providers for the short-term trials. A novel cooperative long-term QoS prediction approach is developed that utilizes past trial experiences of similar consumers using a workload replay technique. We propose a new trial workload generation model that estimates a provider's long-term performance in the absence of past trial experiences. The confidence of the prediction is measured based on the trial experience of the consumer. A set of experiments are conducted based on real-world datasets to evaluate the proposed framework.

% A collaborative filtering based approach is proposed to predict an IaaS provider's long-term performance using past consumers' trial experience. 

% We design a novel approach to map the long-term service requirements into the short-term trials using time series compression techniques. The long-term performance is predicted using a novel time series expansion technique on the short-term trial experiences. 
\end{abstract}

% Note that keywords are not normally used for peerreview papers.
\begin{IEEEkeywords}
Service Selection, Long-term IaaS, Temporal Skyline, and Cooperative Performance Prediction
%\vspace{-3m}
\end{IEEEkeywords}}

% make the title area
\maketitle

% To allow for easy dual compilation without having to reenter the
% abstract/keywords data, the \IEEEtitleabstractindextext text will
% not be used in maketitle, but will appear (i.e., to be "transported")
% here as \IEEEdisplaynontitleabstractindextext when the compsoc 
% or transmag modes are not selected <OR> if conference mode is selected 
% - because all conference papers position the abstract like regular
% papers do.
\IEEEdisplaynontitleabstractindextext
% \IEEEdisplaynontitleabstractindextext has no effect when using
% compsoc or transmag under a non-conference mode.

% For peer review papers, you can put extra information on the cover
% page as needed:
% \ifCLASSOPTIONpeerreview
% \begin{center} \bfseries EDICS Category: 3-BBND \end{center}
% \fi
%
% For peerreview papers, this IEEEtran command inserts a page break and
% creates the second title. It will be ignored for other modes.
\IEEEpeerreviewmaketitle

\IEEEraisesectionheading{\section{Introduction}\label{sec:introduction}}
% Computer Society journal (but not conference!) papers do something unusual
% with the very first section heading (almost always called "Introduction").
% They place it ABOVE the main text! IEEEtran.cls does not automatically do
% this for you, but you can achieve this effect with the provided
% \IEEEraisesectionheading{} command. Note the need to keep any \label that
% is to refer to the section immediately after \section in the above as
% \IEEEraisesectionheading puts \section within a raised box.

% The very first letter is a 2 line initial drop letter followed
% by the rest of the first word in caps (small caps for compsoc).
% 
% form to use if the first word consists of a single letter:
% \IEEEPARstart{A}{demo} file is ....
% 
% form to use if you need the single drop letter followed by
% normal text (unknown if ever used by the IEEE):
% \IEEEPARstart{A}{}demo file is ....
% 
% Some journals put the first two words in caps:
% \IEEEPARstart{T}{his demo} file is ....
% 
% Here we have the typical use of a "T" for an initial drop letter
% and "HIS" in caps to complete the first word.
% \IEEEPARstart{T}{his} demo file is intended to serve as a ``starter file''
% for IEEE Computer Society journal papers produced under \LaTeX\ using
% IEEEtran.cls version 1.8b and later.

Cloud computing has become a pivotal technology of choice for large and medium business organizations to establish and manage their IT infrastructure \cite{chaisiri2012optimization}. Large organizations such as governments, financial institutions, and academic institutions tend to utilize cloud services on a \textit{long-term} basis for economic reasons \cite{van2014optimizing}. Cloud providers such as Amazon, Google, and Microsoft promote long-term subscriptions by advertising long-term (e.g., 1 to 3 years) services at lower prices \cite{mazzucco2011reserved}. Cloud services are generally categorized as Infrastructure-as-a-Service (IaaS), Platform-as-a-Service (PaaS), and Software-as-Service (SaaS).

\textit{IaaS provider selection} for a long-term period is a topical research challenge \cite{scheuner2018estimating,ye2014long}. Selecting the right IaaS provider is an important \textit{business decision} for cloud consumers \cite{davatz2017approach}. IaaS providers are selected based on a consumer's long-term \textit{functional} and \textit{non-functional} requirements \cite{scheuner2018estimating}. Functional requirements are set based on the purpose of the service such as computing, data storing, and networking.  Non-functional requirements are often expressed in terms of Quality of Service (QoS) attributes such as availability, response time, and throughput. QoS attributes help a consumer to select the \textit{best performing} services from a large number of functionally similar services. The QoS-aware service selection is therefore defined as the similarity matching between a consumer's long-term QoS requirements and the expected long-term performance of IaaS services \cite{ye2014long}. %An optimal service has the highest similarity matching with the consumer's long-term QoS requirement than any other candidate services. We use the word ``optimal'' and ``best'' interchangeably.

Selecting an appropriate IaaS provider is challenging without the detailed information of a provider's long-term QoS performance. IaaS providers are typically reluctant to divulge the \textit{detailed} and \textit{complete} information about their long-term QoS management policies in the dynamic multi-tenant environment \cite{dou2013hiresome}. The key reasons for such behavior are market competition, business secrecy, and conflict of interests \cite{vicentini2018machine}. For example, Amazon does not disclose the actual throughput information of its vCPUs in the advertisements\footnote{https://aws.amazon.com/ec2/instance-types/}. A consumer who has CPU-intensive workloads may find it challenging to select Amazon for a long-term period with such limited performance information \cite{scheuner2018estimating}.

To the best of our knowledge, existing approaches mainly focus on two aspects for the IaaS provider selection with the \textit{incomplete} information: a) relying on IaaS advertisements, and b) gathering information from trial experiences \cite{wang2018testing}. IaaS advertisements typically contain limited QoS information. For instance, Each vCPU may be a thread of an Intel Xeon core, an AMD EPYC core, or AWS Graviton processor according to EC2 advertisements\footnotemark[1]. Estimating the performance of the vCPU is difficult from such limited information \cite{scheuner2018estimating}.  Most providers do not differentiate between long-term and short-term services in terms of their performance. Providers often fail to offer their advertised QoS performance. Therefore, relying on only IaaS advertisements is not sufficient for long-term selection.

IaaS providers offer \textit{free trials} encouraging potential consumers to test their application workloads in the cloud. For instance, Microsoft Azure offers a one-month trial period for a limited number of services to its potential consumers\footnote{https://azure.microsoft.com/en-us/free/}. IaaS consumers may get a first-hand experience before subscribing to a service for the long-term period. The trial experience of consumers may have a considerable impact on the IaaS provider selection process \cite{zhu2014investigating}.

Application benchmarks and micro-benchmarks are usually utilized in the trial periods to discover a provider's performance in terms of various QoS attributes such as CPU speed, disk read/write latency, and network bandwidth \cite{scheuner2018estimating}. Synthetic workloads are generated for the representative applications to perform \textit{stress testing} on the providers. The results of the tests are used to compare different providers. These approaches mainly focus on the \textit{short-term selection} and do not consider the long-term QoS performance variability of the providers. In reality, the QoS performance of a provider varies over the long-term period due to the \textit{multi-tenant} nature of the cloud environment \cite{iosup2014iaas}. For example, a provider may show very good performance in the Christmas period, when there are very less number of active consumers.

\textit{We aim to leverage the free short-term trial experience of consumers for the long-term IaaS provider selection.} We identify the following key challenges to select IaaS providers for a long-term period based on short-term trial experience:

\begin{itemize}[itemsep=0ex, leftmargin=2ex]
    \item \textit{Number of Candidate IaaS Providers}: A large number of IaaS providers may satisfy a consumer's long-term requirements. The performance of these providers may vary considerably depending on their business strategies and infrastructures. Performing trial with every eligible IaaS is \textit{practically} infeasible.
    
    \item\textit{Temporal Restrictions}: Trial periods are generally offered for a \textit{short-term period}. A consumer cannot test its entire long-term workloads in the short-term trial period. Some IaaS providers (e.g., Amazon) offers long-term trials for several services\footnote{https://aws.amazon.com/free/}. A consumer may not be able to \textit{wait such a long time} to make the selection.

    \item \textit{Performance Variability}: It is challenging to predict the long-term performance from a short trial without the performance variability information \cite{wang2018testing}. Most IaaS providers usually operate in multi-tenant environments. IaaS performance is highly time-dependent \cite{iosup2014iaas}. The performance discovered in a one month trial period may not \textit{reflect} the actual performance of a provider for the rest of the year.

\end{itemize}

Trial experience of a consumer may depend on several factors such as trial workloads, the time of the trial, and the provider's QoS management policy \cite{scheuner2018estimating}. The experience from an \textit{unplanned} short trial may not provide the complete information required for the long-term selection.

We propose a novel framework that utilizes a consumer's short-term trial experience to select the closest match IaaS provider according to the consumer's long-term QoS requirements. We incorporate the experience of past trial users to predict providers' long-term performance. Experience of past trial users may not be applicable directly to predict IaaS performance for a new consumer. The reason is that the workloads of the past trial users are most likely to be \textit{different} from the new consumer's workloads. We propose a cooperative long-term QoS prediction approach based on the \textit{workload similarity} of the past trial users. The performance of the provider may change over the long-term period. The proposed framework measures the confidence of the prediction based on the trial experience of the new consumer. Our contributions are summarized as follows:

\begin{itemize}[itemsep=0ex, leftmargin=2ex]
    \item A skyline-based filtering method to select candidate IaaS providers to perform free trials. 
    \item A Cooperative long-term QoS performance prediction approach using the experience of past trial users. The confidence of the trial is measured based on the trial experience of the new consumer.
    \item A long-term QoS performance prediction approach using a trial workload generation model when no similar trial users are available.

    \item A QoS-aware selection method to choose the ``closest match" IaaS provider using multidimensional time series similarity measure between the consumer performance requirements and the predicted performance of providers.
\end{itemize}

%\vspace{-2mm}
\section{Motivation Scenario}

\begin{figure}
    \centering
    \includegraphics[width=0.45\textwidth]{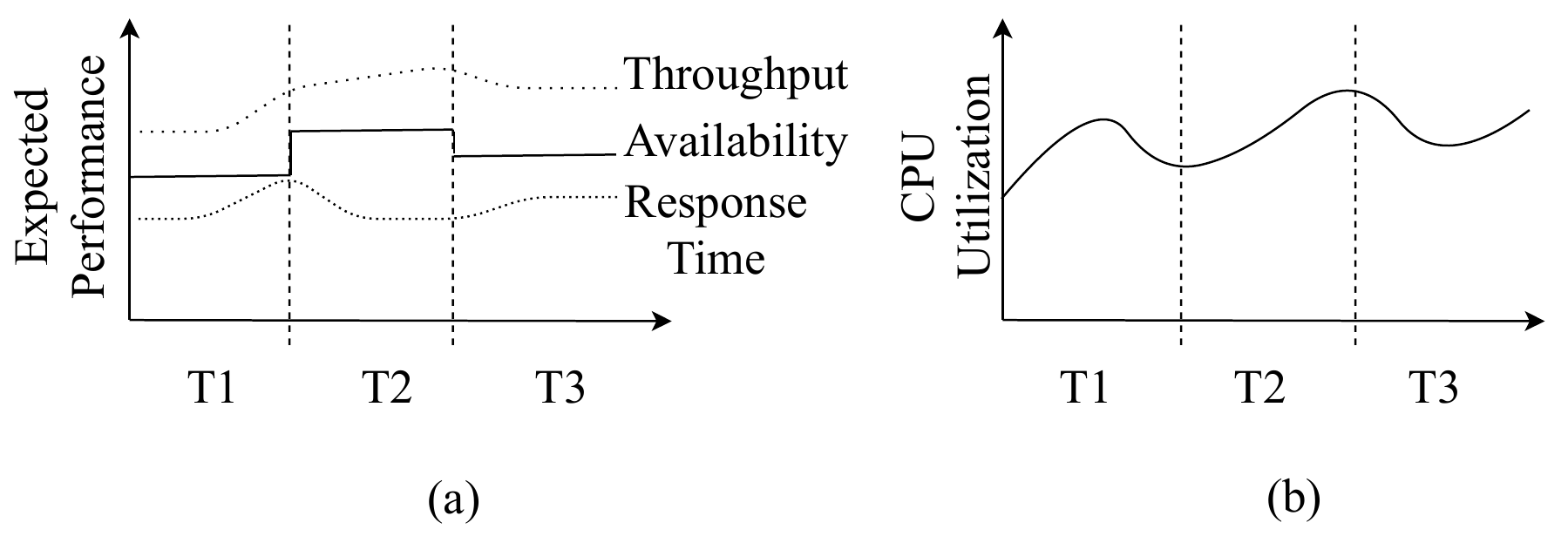}
    \caption{University's long-term requirements (a) Expected QoS performance (b) Long-term CPU workloads}
    \label{fig:con_req}
    \vspace{-2mm}
\end{figure}
%\vspace{-2mm}
% \begin{figure}
%     \centerline{
%     \subfloat[]{\includegraphics[width=.24\textwidth]{images/conReq.pdf}}
%     \hfil
%     \subfloat[]{\includegraphics[width=.24\textwidth]{images/conReq1.pdf}}
%     }
%     \caption{University's long-term requirements (a) Expected QoS performance (b) Long-term CPU workloads}
%     \label{fig:con_req}
% \end{figure}

Let us assume that a university wants to lease some general-purpose VMs for one year. Each VM requires at least 2 vCPU and 6 GB memory. The university has a minimum expected QoS performance on availability, response time, and throughput. The QoS requirements vary based on the university's seasonal demands. For example, the university may need high throughput during the examination period, and low throughput in the Christmas period. Figure \ref{fig:con_req}(a) depicts the consumer's expected availability on three different periods T1, T2, and T3 in a year. We assume that the university's workloads are \textit{deterministic}, i.e., the university has estimated its future workloads for one year based on history. Figure \ref{fig:con_req}(b) shows the consumer's CPU workloads for $T1$, $T2$, and $T3$ period.

\begin{figure}
    \centering
    \includegraphics[width=0.45\textwidth]{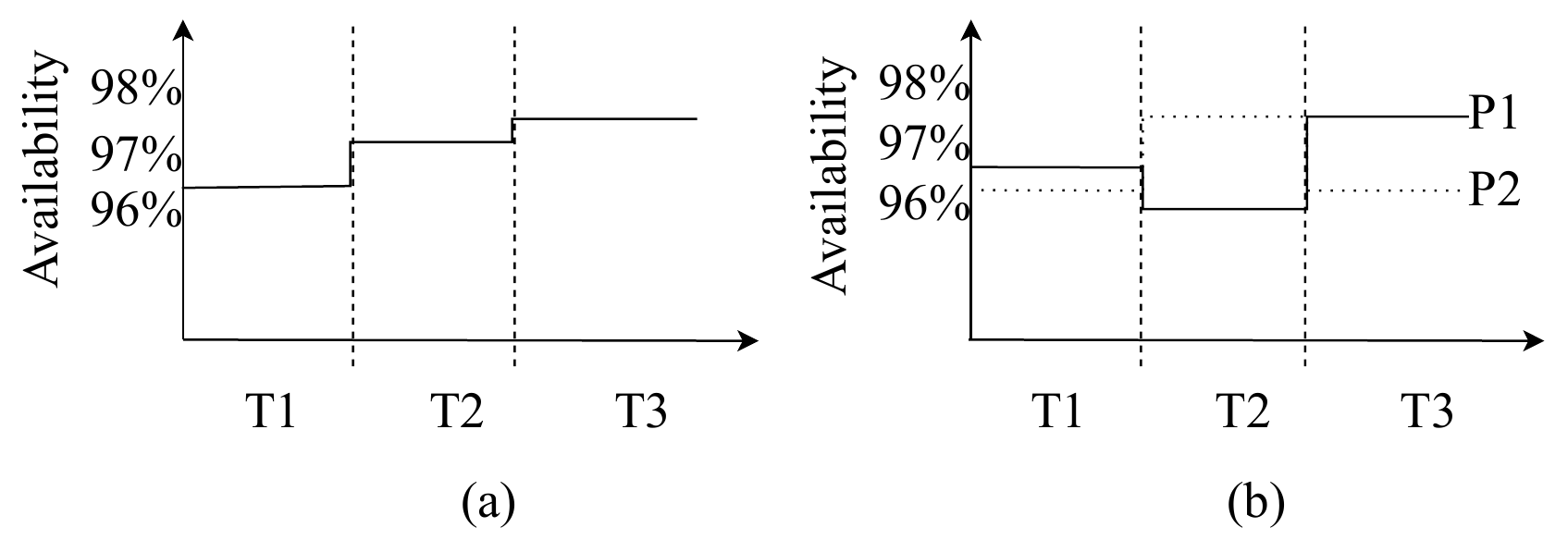}
    \caption{(a) Expected availability (b) Advertised availability}
    \vspace{-6mm}
    \label{fig:advertisement}
\end{figure}

Several IaaS providers in the cloud market may advertise the required type of VM to the university. Let us assume that $P1$ and $P2$ are two such providers. Both providers advertise availability information for the VMs in their IaaS advertisements. The response time and throughput information are not available in the advertisements. Figure \ref{fig:advertisement}(a) shows the consumer's expected availability. Figure \ref{fig:advertisement}(b) shows the advertised availability of two providers. The consumer will select $P2$ if it selects based on availability. This selection may not be a good decision as it does not consider the consumer's expected throughput and response time. 

Let us assume that each provider offers a trial to the university.  The university performs trials on the $T2$ period and observes the throughput and the response time of the IaaS service. Note that, the university cannot test its entire workloads in the $T2$ period. Let us assume that the university performs free trial only using the workload of the $T2$ period (Figure \ref{fig:con_req}(a)). If the university makes the selection considering only the observed performance in the trial period, it may not be the right decision. For example, the university may observe that $P1$ offers better throughput and response time. Selecting $P1$ may be a wrong decision as $P2$ performs better in $T1$ and $T2$ period.

Let us assume that \textit{the consumer has access to the experiences and workloads of past trial users}. Workloads of trial users may be different from the university's workloads. We identify the following cases:

\begin{itemize}[itemsep=0ex, leftmargin=2ex]
    \item \textit{Similar Trial Users}: The experience of past trial users is available where the users have similar workloads to the university. These users could be universities, colleges, banks, and other large organizations. The university performs a trial using its workloads of $T2$ period and observes the throughput and response time of the providers. The university utilizes the experience of the trial users to predict the performance of each provider for its workloads in $T1$ and $T3$ period. The confidence of the prediction needs to be measured as the prediction is made based on historical information. Each provider's current performance may be improved or degraded than the experiences in the past.
    
    \item \textit{Dissimilar Trial Users}: In this scenario, we assume that there are no trial users available who have similar workloads to the university. If the existing  trial users include SaaS providers and video content providers, they are most likely to have different workloads than the university. In such a case, the university can not completely rely on the past experiences. The university needs to perform the trial efficiently to understand the providers' performance behavior for its long-term workloads.  
\end{itemize}

\textit{The proposed selection framework considers both cases for long-term selection}. First, a filtering method is applied that selects candidate IaaS providers for the trial periods. Next, the proposed framework predicts the long-term QoS performance of each provider by leveraging the trial periods for similar trial users and dissimilar trial users. An IaaS provider is then selected based on the predicted QoS performance and the consumer's expected QoS performance from the candidate providers. %We introduce and describe a long-term IaaS provider selection framework in the following sections. 

%\vspace{-2mm}
\section{Related Work}

%benchmarking

Several approaches are proposed to compare IaaS providers based on standard \textit{benchmarks} \cite{iosup2014iaas}. A common approach is to utilize existing benchmarks to compare the performance of different providers. A novel benchmarking approach is proposed in \cite{binnig2009weather}. The paper argues that existing benchmarks for computer systems are not suitable. The paper identifies some general requirements and constraints for TPC benchmarks, which are not suitable for clouds. The paper calls for a new cloud benchmarking approach that should consider scalability, cost, and fault-tolerance during the average performance measurement.  A systematic comparator of cloud performance called CloudComp is proposed to help cloud consumers choosing a provider that meets its requirements \cite{li2010cloudcmp}. CloudComp is applied to four popular cloud providers and finds that cloud service performance and costs vary widely among cloud providers. 

%application testing
Various experiments are conducted to understand the performance of IaaS providers for \textit{different types of applications}. IaaS cloud services are tested with diverse application workloads. The performance behavior of Amazon EC2 small instances for service-oriented application is studied in \cite{dejun2010ec2}. The proposed study benchmarks virtual instances by generating different types of workload patterns and analyzes the performance in terms of mean response time. A generator approach is proposed to automate performance testing in IaaS cloud \cite{jayasinghe2012expertus}. The proposed approach utilizes a template-driven code generation method to test different applications on IaaS clouds. These approaches do not consider the long-term performance variability. Therefore,\textit{ these works are not applicable for the long-term performance discovery.}

The \textit{experience of existing users} are utilized to predict QoS ranking of cloud providers \cite{wang2019qos,yang2018location, tang2016collaborative}. A collaborative filtering (CF) approach is proposed to predict QoS values based on historical QoS information provided by existing users \cite{wang2019qos}. The proposed approach finds similar users based on a user's QoS requirements and utilizes similar users' experience to predict personalized QoS ranking. A location-aware CF approach is proposed to predict missing QoS parameters for web service recommendation \cite{yang2018location}. This study suggests that the location of a user has a remarkable impact on the value of QoS attributes such as availability, response time, and throughput. The proposed location-aware CF approach improves the performance of recommendation significantly by incorporating location information of both users and services in existing similarity measurement approaches of CF. These approaches mainly focus on short-term prediction and do not consider the providers' performance variability over a long period of time. Several studies has focused on time-aware QoS prediction approaches \cite{qi2019time,hu2014time}.  A time series forecasting (TSF) based approach is studied to predict the QoS performance of cloud providers \cite{zadeh2010qos}. The proposed approach uses Neural Networks (NN) for time series forecasting to conduct experiments. The experiments show promising results for using TFS to predict QoS performance. These approaches are mainly focused on web services and do not take the consumer's workload into consideration during the QoS prediction. Therefore, \textit{these approaches are not directly applicable to our work}. 

A QoS prediction model is proposed using naive Bayesian classifiers in \cite{al2018performance}. The proposed model uses historical information perceived by end-users to predict different performance metrics of cloud-based on different configurations of VMs. A new performance prediction method is proposed in \cite{scheuner2018estimating}. The proposed approach build classifiers based on application and micro-benchmark results to estimate cloud application performance on VMs. These approaches do not consider the long-term performance variability. As a result, these approaches cannot be applied directly for the long-term performance discovery.

Testing a consumer's long-term workloads in the cloud may be cumbersome and \textit{error-prone}. The performance of the same application in two different clouds may vary significantly. Most existing studies conduct experiments to measure \textit{short-term performance}. Existing performance monitoring and testing approaches are not directly applicable to the long-term selection. Availability of historical datasets with detailed information is limited in the public domain due to the privacy issue. \textit{Sharing trial experience is less concerning to the cloud consumers} \cite{swan2012crowdsourced}. There are forum and websites such as Geekbench\footnote{https://www.geekbench.com/} where consumers share their trial experience. We leverage the trial experience of past consumers to predict future QoS performance of a provider. The long-term selection is performed for a new consumer based on its \textit{trial confidence} and the long-term requirements.

\vspace{-1mm}
\section{IaaS Selection Framework}

We formulate the long-term IaaS provider selection using the following definitions and notations in Table \ref{tab1}.

\begin{itemize}[itemsep=0ex, leftmargin=2ex]
\item \textit{Consumer}: A consumer is a new IaaS Consumer who requires Virtual Machines (VMs) for a long-term period.
\item \textit{Functional Requirements}: Functional requirements are defined in terms of different types of VM configurations (i.e., the number of vCPU and memory units) and the number of VMs over the long-term period. 
\item \textit{Long-term Workloads}: The workload of a consumer is represented as the requested number of resource units such as CPU and memory over the long-term period.
\item \textit{QoS Requirements}: QoS requirements of a consumer is a set of QoS parameters and their minimum or average expected values for a long-term period. 
\item \textit{Provider}: A provider is an IaaS provider who provisions VMs for a long-term period.
\item \textit{QoS Advertisement}: A QoS advertisement is a set of QoS parameters for a VM and the values of the QoS parameters over the long-term period.
\item \textit{Trial Periods}: A consumer can use some services with restricted conditions for a short time for free. 

\end{itemize}

Let us assume that a consumer \textit{requires} a set of general-purpose VMs of a particular configuration over a long-term period. The consumer has \textit{variable workloads} over the long-term period. We assume that the workloads are \textit{deterministic} i.e., the consumer has full knowledge of its workload distribution per VM over the long-term period. The workload of the consumer can be defined as the required amount of resources (e.g., CPU time, and Memory size) at a particular period. \textit{We consider the workload as a combined resource requirement at a particular time and denote it as $W$.} 

We represent the consumer's workloads and QoS requirements for the long-term period using \textit{time series groups} (TSGs). We denote the total service usage time as \(T\). The TSG of QoS requirements is defined as \(Q_C = \{q_{c1},q_{c2},...,q_{cl}\} \), where \(cl\) is the number of QoS parameters in \(Q_C\) and \( q_{cl} = \{(x_n,t_n)|n=1,2,3,...., T\} \), where \(x_n\) is the value of \(q_{cl}\) at the time of \(t_n\). We define the time series of workloads per VM as \(W = \{(w_n,t_n)|n = 1,2,.., T\}\), where \(w_n\) is the workload per VM at time \(t_n\). Here, n denotes a timestamp.

Let us assume that there are \(N\) number of IaaS providers who can fulfill the functional requirements of the consumer. The set of the providers is denoted as \(P= \{P_1,P_2,...P_N\}\). The QoS performance of VMs varies from one provider to another provider over the long-term period. Each provider advertises long-term QoS properties in TSG for its VMs. The QoS advertisement of the provider \(P_i\) is denoted as \(A_i = \{a_{i1},a_{i2},....a_{ik}\} \), where \(ik\) is the number of QoS parameters in \(A_{i}\). We assume \(ik<l\), i.e., the number of QoS parameters in the advertisements is always less than the number of the QoS parameters in the consumer requirements. The time series of each QoS parameter of provider \(P_i\) in the advertisement is denoted as \( a_{ik} = \{(y_{im},t_{im})|im=1,2,3,...., T\} \) where \(y_{im}\) is the value of \(a_{ik}\) at the time of \(t_{im}\). Here, \(im\) is the timestamp of the advertisement of the provider \(P_i\). We assume that each provider advertises with the same interval, i.e., \(im\) is \textit{same} for all providers.

The consumer requires to predict long-term QoS performance of the providers to make an informed selection. The advertisements do not provide enough information (i.e., \(ik<l\)). We denote the predicted QoS performance of the provider $P_i$ as $Q_i=\{q_{i1},q_{i2},...,q_{l}\}$ where \(l\) is the number of QoS parameters in \(Q_{i}\). The consumer requires to select the a provider based on the predicted QoS performance that closely matches its expected QoS performance. Given the consumer's QoS expectations $Q_C$ and a provider's predicted QoS performance $Q_i$, we use a predefined TSG distance measuring function $distance(Q_C,Q_i)$ to find the most similar provider $P_s$ using the following equation:

\small

\begin{equation}
P_s = \text{argmin}_{i=1}^l (distance(Q_C,Q_i))
\label{eqn:problem}
\end{equation}

\normalsize

\begin{table}[!t]
\caption{Notations and Descriptions}
 \begin{tabular}{ l l } 
 \hline
 Notation & Description \\ [0.5ex] 
 \hline\hline
 \(T\) & Required provisioning time \\
 \(W\) & Workload requirements in time series \\ 
 \(Q_C\) & Set of QoS Requirements of Consumers\\
 \(l\) & Number of QoS parameters in \(Q_C\) \\
 \(q_{ci}\) & The time series of the QoS parameter \(q_{ci}\) \\
 \(t_n\) & A timestamp in \(T\) where \(n=1,2,3,...T\) \\
 \(x_n\) & The value of \(q_(ci)\) at the time period \(t_n\) \\
 
 \(N\) & The number of IaaS providers who can\\
         & satisfy the functional requirements of the consumer\\ 
 \(P\) & A set of IaaS providers \\
 \(A_{i}\) & The QoS advertisement of the provider \(P_i\) \\
 \hline
\end{tabular}

\label{tab1}
\vspace{-4mm}
\end{table}

\begin{figure}
    \centering
    \includegraphics[width=0.4\textwidth]{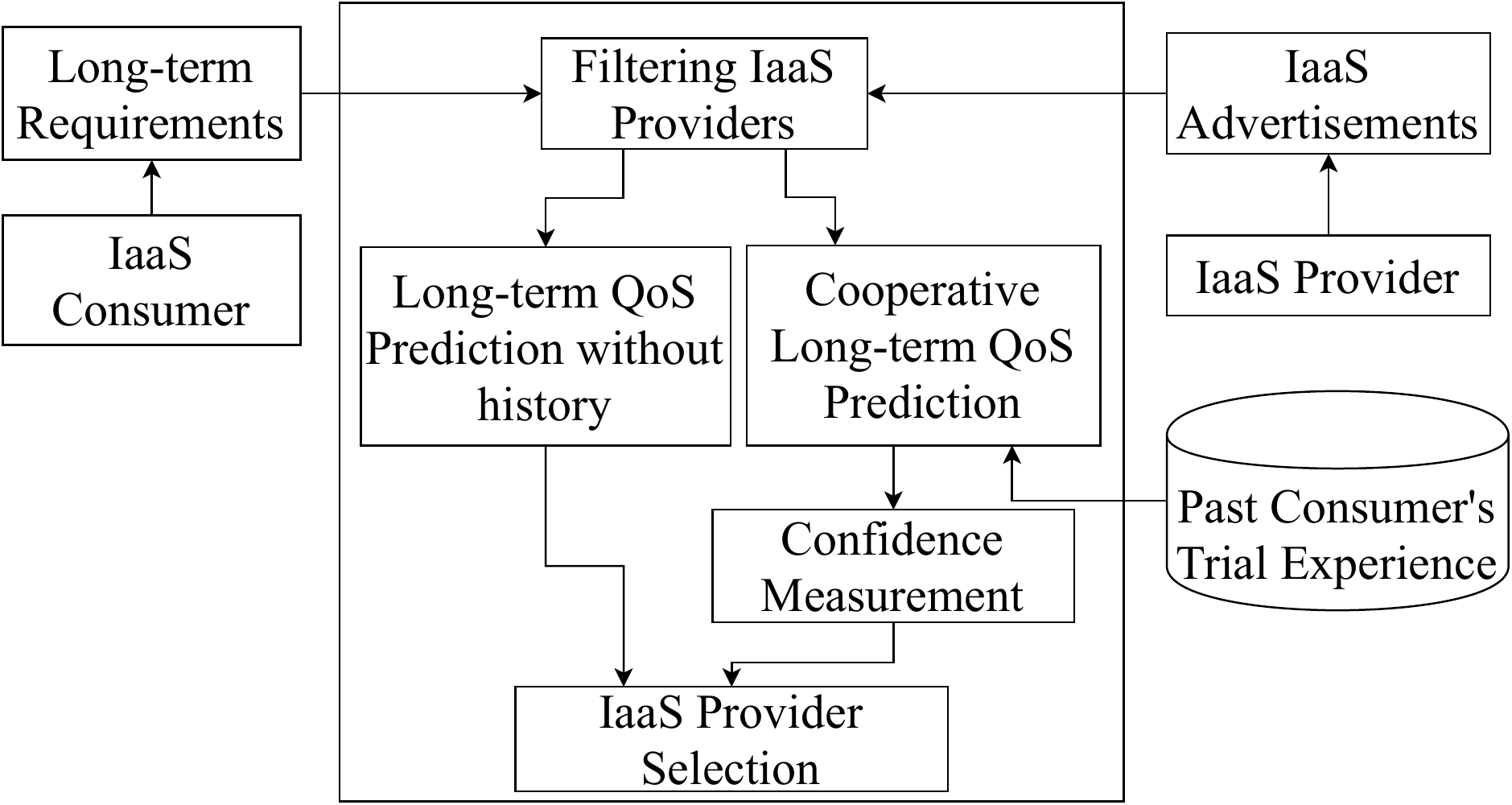}
    \caption{Long-term IaaS provider selection framework}
    \label{fig:framework}
    \vspace{-5mm}
\end{figure}

Figure \ref{fig:framework} shows the proposed long-term IaaS provider selection framework. The framework requires a consumer's long-term requirements, i.e., expected QoS performance and workloads as the \textit{input}. The other inputs to the framework are the long-term advertisements of a set of providers who can fulfill the functional requirements of the consumer. The framework has the following four modules:

\begin{itemize}[itemsep=0ex, leftmargin=2ex]
    \item \textit{Filtering IaaS Providers}: 
    %The number of providers who may fulfill the consumer's functional requirements may be large. The consumer may not be able to perform a trial on each provider. 
    A filtering method is devised to reduces the search space based on the similarity between the consumer requirements and the QoS advertisements of the providers. Finding such providers is a \textit{multi-criteria decision-making problem}. Each provider may have advertisements that are the best for a particular QoS parameter at a particular time period. There may be \textit{no clear winner} who provides the best advertisements for all QoS parameters over the entire period. We therefore incorporate a \textit{temporal skyline} \cite{wang2013dominant} to solve the multi-criteria decision-making problem in the filtering method. 
    \item \textit{Cooperative Long-term QoS Prediction (CLQP)}: A consumer's long-term workloads may not be tested in a short trial. Some parts of the workloads may be tested directly in the trial period while the performance of the other parts can be inferred from past trial users. A cooperative QoS prediction approach is applied based on the \textit{experience of trial users} using a ``workload similarity'' measure technique. A confidence measurement technique is devised for the predicted performance using the trial experience.

    \item \textit{Long-term QoS Prediction without History (LQP-short)}: When there is no similar past trial user, the proposed framework generates trial workloads by mapping the consumer's long-term workloads into the short-term trial periods to discover the performances. The trial workloads are generated using \textit{time series compression techniques}. 
    \item \textit{QoS-aware Long-term IaaS Provider Selection (QLIS)}: The ``closest match'' provider is selected where the predicted performances closely match with the consumer's requirements with the \textit{highest trial confidence} when similar past trial users are available (CLQP). Therefore, the proposed framework does not require an \textit{exact similarity measure} between the consumer requirements and a provider's performance. In the absence of similar trial users, the provider is selected based on the LQP-short. 
\end{itemize}

\vspace{-2mm}
\section{Filtering IaaS Providers}

We develop a filtering method to reduce the number of candidate IaaS providers for the trial. The filtering process can be modeled as a single criterion or multiple criteria decision making depending on the consumer's requirements. 

\vspace{-1mm}

\subsection{Filtering IaaS Providers based on Single Criterion}

If a consumer's QoS requirement ($Q_c$) contains only one QoS parameter, i.e., $|Q_c|=1$, the filtering method can be modeled based on single criterion decision making. For example, when a consumer only cares about the response time of the VMs, the candidate providers can be selected based on the response time without considering other QoS attributes. In such a case, we compare the consumer's long-term QoS requirement with each provider using time series similarity matching techniques. A well-known time series similarity matching technique is the Mean Absolute Error (MAE) distance \cite{tang2016collaborative}. MAE is a fast, effective, and easy to implement technique where similarity is measured between each corresponding timestamp of two time series. The Equation \ref{eqn:mea} computes the similarity between the consumer's long-term QoS requirement $q_c$ and a provider's advertisement $(a_p)$ for a single QoS parameter:

\small
\vspace{-2mm}
\begin{equation}
    \text{MAE } (q_c,a_p) =  \frac{1}{n}{\sum_{t=1..n} \mid q_c^t-a_p^t \mid}
    \label{eqn:mea}
\end{equation}

\normalsize

In Equation \ref{eqn:mea}, $n$ is the number of timestamps, $q_c^t$ is the value of QoS parameter $q_c$ at time $t$. A number of candidate providers should be selected based on the measured distance. The candidate providers can be selected using the Top-K approach \cite{zheng2012qos}. The Top-K approach selects the best $K$ number of candidates who have the minimum RMSE distance from the consumer's QoS expectation. If the number of selected candidate providers are too small or too large, $K$ can be adjusted to reduce or to increase the number of candidate providers for the trial period.

\subsection{Filtering IaaS Providers based on Multiple Criteria}
When the consumer's QoS requirements ($Q_c$) contains more than one QoS parameter, i.e., $|Q_c|>1$, we can model the filtering method as a multi-criteria decision-making problem. Multi-criteria decision-making approaches are applicable when the consumer's selection depends on multiple criteria such as price, throughput, and response time. In such a case, the consumer's QoS requirements and a provider's QoS advertisements should be compared for each QoS parameter. The main issue of filtering based on multi-criteria is having \textit{pareto-optimality} or incomparable providers. For instance, a provider may advertise VMs at the lowest price where another provider may advertise VMs with the highest performance. In such a case, there is no \textit{clear winner} as both of them are best on different criteria. 

\subsubsection{Filtering based on Utility Function}
There are several approaches to filter candidate providers based on multiple criteria. Some general approaches are utility function, conditional preference, and skyline \cite{jiang2009online}. The \textit{utility function} computes a score of each provider based on the consumer's preference for each attribute. A consumer needs to assign weights on each QoS attributes based on its preference. For example, a consumer may put the highest weight on the price attribute as the price may be the most important attribute. The consumer may set lower weights to the lesser important attributes.  The utility function calculates a score for each provider based on the weights of the attributes and the value of the attributes. The utility function computes the score using the following equation: 

\small

\begin{equation}
    \text{Score} (P) =  \sum_{q_c \in Q_C, a_p \in A_p} W_{q} \times \text{MAE}(q_c,a_p) 
\end{equation}

\normalsize

where $W_q$ is the weight of the QoS attribute $q$ assigned by the consumer, $Q_c$ is the TSG of consumer's QoS requirements, $A_p$ is a provider's QoS advertisements. A multi-criteria decision making problem is then transformed into a single-criterion decision making problem using the utility function. We can apply the TOP-K method using the utility scores to filter the providers.

\subsubsection{Filtering IaaS Providers based on IaaS Skyline}

The utility function based filtering method has two major \textit{drawbacks}. First, it requires a consumer to assign weights on the QoS attributes. Second, if a consumer's preferences change the utility function should be updated. The skyline algorithm is a well-known alternative to the utility function \cite{wang2013dominant}. A skyline does not require a consumer to assign weights to the QoS attributes. It includes all of the outputs that can be generated by a utility function \cite{jiang2009online}. The skyline concept is adopted from the real-world skyline where the most dominating items are displayed in a skyline. For example, A city skyline consists of the tallest, widest, or closest building to the viewer. Similarly, a skyline filtering method consists of the provider who advertises the best value of at least one of the QoS attributes.

Let us assume that there are \(N\) IaaS providers who can fulfill the consumer's functional requirements. The number of such IaaS providers is not expected to very large given the number of top IaaS providers is no more than 100 \footnote{\url{https://stackify.com/top-iaas-providers/}}. Therefore, we consider the value of $N$ here is arbitrary. We need to find a set of candidate providers \(P'\) where \(P' \subset P \) and \(|P'| < |P|\), i.e., the number of selected providers are less than the number of existing providers. Let us assume that \(|P'|=M\) i.e., number of selected IaaS providers is M.  The set of QoS advertisements for all IaaS providers is denoted as \(A = \{A_1,A_2,A_3,...,A_N\}\) where \(A_i\) is the advertisement of provider \(P_i\). We need to select \(M\) number of IaaS providers from the existing \(N\) IaaS providers based on the advertisements \(A\) and the consumer requirements \(Q_C\). We need to find a set \(A'\) where \( A' \subset A \) and \(A'\) contains the closest match information according to \(Q_C\). The advertisement contains more than one QoS parameter to compare for ranking the providers. Therefore, selecting the candidates is a multi-criteria selection problem.

We utilize a \textit{temporal skyline} to solve the selection of candidate providers over multiple QoS criteria. The \textit{skyline query} is an effective approach to select a set of data points that are better than other data points in a large dataset \cite{jiang2009online}. The temporal skyline is a time-series version of the skyline that selects a set of time series that are better than any other time series. We propose a temporal skyline approach to select a set of candidate providers who are better than the other set of providers over time \(T\). In the following subsections, we define the IaaS skyline and Temporal IaaS skyline to select the candidate providers for free trials. 

\subsubsection{IaaS Skyline}

\begin{figure}
    \centerline{
        \subfloat[]{\includegraphics[width=0.24\textwidth]{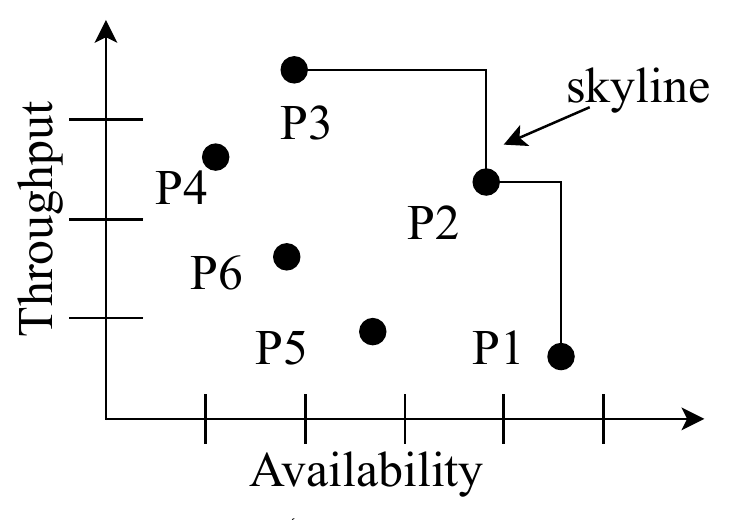} \label{skypoint}}
        \hfill
        \subfloat[]{\includegraphics[width=0.24\textwidth]{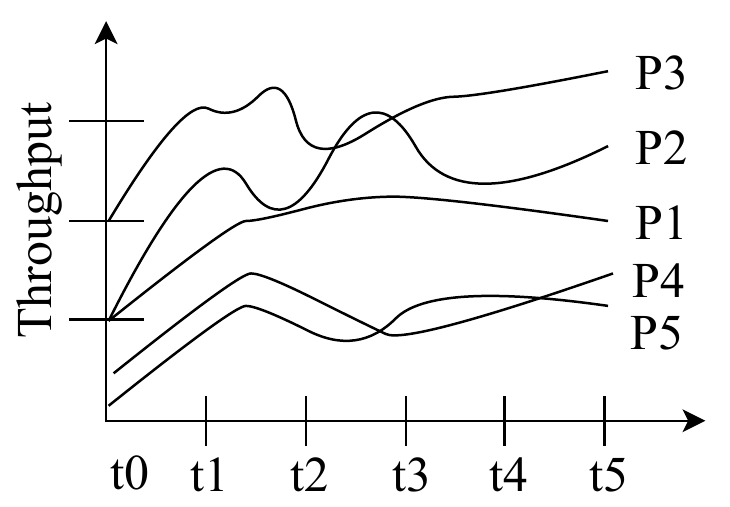} \label{skytime}}
      
    }
 
\caption{Filtering with skyline (a) IaaS skyline (b) Temporal IaaS skyline}
\label{fig:skypoints}
\vspace{-4mm}
\end{figure}

A provider \(P_i\) dominates another provider \(P_j\) if \(P_i\) offers equal or better QoS advertisements than \(P_i\) for all QoS parameters and \(P_i\) offers better advertisement than \(P_j\) for at least one QoS parameters. A skyline of providers contains the providers that are \textit{pareto-optimal} and not dominated by any other providers for every QoS parameters. Figure \ref{fig:skypoints}(a) shows an example of the skyline of providers. Here, each provider offers two QoS parameters, i.e., throughput and availability. \(P_3\) offers higher throughput and availability time than \(P_4\) and $P_6$ thus \(P_3\) dominates \(P_4\) and \(P_6\). Similarly, \(P_2\) dominates \(P_5\), \(P_6\), and $P_4$. The skyline of the providers contpains \(P_1\), \(P_2\), and $P_3$. 

We assume that the consumer has a set of QoS parameters that are more important or \textit{dominant} than other QoS parameters denoted as \(Q_D\) where \(Q_D \subset Q_C\). For example, a consumer may consider that the throughput and availability of a VM are more important than any other QoS parameters. The weight of each dominant QoS parameter is considered equal. A provider who does not provide throughput and availability information will not be considered for the trial. The size of \(Q_D\) can vary according to consumer preferences.

\begin{definition}{\textit{Dominant Provider}}. A provider \(P_i\) dominates another provider \(P_j\), denoted as \(P_i \succ P_j\), if \(P_i\) provides as good or better advertisements for all dominant QoS parameters in \(Q_D\) i.e., $\forall q \in Q_D : P_i \succeq P_j$ and $\exists q' \in Q_D : P_i \succ P_j$
\end{definition}

\begin{definition}{\textit{IaaS Skyline}}. The IaaS skyline of a set of IaaS Providers $P$, denoted as $SK_P$, is a subset of providers that are not dominated by any other providers i.e., $SK_P=\{p \in P | \neg \exists p' \in P : p' \succ p\}$
\end{definition}

\subsubsection{Temporal  IaaS Skyline}

We represent the long-term QoS advertisements using time series where the value of each QoS parameter may change over time. We utilize the temporal skyline introduced in \cite{jiang2009online}, where QoS parameters can be represented as time series . 

\begin{definition}{\textit{Dominant QoS Time Series}}. A QoS time series $Q_i$ dominates another QoS time series $Q_j$ in time $T$, denoted as $Q_i \succ Q_j$, if $\forall t \in T, Q_i \succeq Q_j $ and $\exists t' \in T, Q_i \succ Q_j$. 

\end{definition}

\begin{definition}{\textit{Temporal QoS Skyline}}.  The temporal QoS skyline of a set of QoS time series $Q$, denoted as $ST_Q$, is a subset of QoS time series that are not dominated by any other QoS time series for all timestamp $t$,  i.e., $ST_Q=\{q \in Q | \neg \exists q' \in Q : q' \succ q\}$
\end{definition}

Figure \ref{fig:skypoints}(b) shows a set of ``throughput" time series of providers $P1$, $P2$, $P3$, $P4$, and $P_5$. $P3$ and $P2$ dominates all other providers for each timestamps, i.e., $t0$ to $t5$. $P3$ dominates $P2$ at $t2$ and $p2$ dominates $p3$ at $t3$. Therefore, $p3$ and $p2$ do not dominate each other and are included in the temporal QoS skyline for the interval from $t2$ and $t3$. The temporal skyline shown in figure \ref{fig:skypoints} considers only a single QoS parameter. In reality, the number of dominant QoS parameters $Q_D$ in the advertisements may be more than one. we need a multiple time series approach for the skyline. We define a multiple time series (MTS)-based \textit{IaaS skyline} and \textit{Temporal QoS Skyline} as follows:

\begin{definition}{\textit{MTS-based Dominant IaaS Provider}}. An IaaS provider $P$ that has $Q_D$ dominant QoS time series, dominates another IaaS provider $P'$, denoted as $P \succ_{mts} P'$ if (1) $\forall t \in T, P \succeq P' $ and $ \exists t' \in T, P \succ P' $ and (2) $\forall q \in Q_D, P \succeq P' $ and $ \exists q' \in Q_D, P \succ P'$

\end{definition}

\begin{definition}{\textit{MTS-based IaaS Skyline}}. The IaaS skyline of a set of IaaS Providers $P$, denoted as $MTS_P$, is a subset of providers that are not dominated by any other providers, i.e., $MTS_P=\{p \in P | \neg \exists p' \in P : p' \succ_{mts} p\}$
\end{definition}

We assume that each provider's advertisement contains at least one dominant QoS parameter. Hence, the MTS-based IaaS skyline includes the providers that are not dominated by any other providers for all timestamp in $T$ and all QoS parameters in $Q_D$. There are several ways of computing skylines in the existing literature \cite{wang2013dominant}. Our focus is on how to apply skyline to select candidate IaaS providers rather than computing skyline efficiently. We use a nested loop (NL) algorithm  \cite{jiang2009online}. Algorithm \ref{alg:sky} illustrates the NL algorithm to compute IaaS skyline.

\begin{algorithm}
    \caption{Computing Skyline using NL Algorithm}\label{alg:workload}
    \begin{algorithmic}[1]
       
        \State \textbf{Input: }$L$, $A$,
        \State \textbf{Output: }$MTS_P$ 
        \\
        $S \leftarrow L$;
        
        \For{$P \in S$}
            \For{$P_i \in S$}
                \If{$P_i \succ_{mts} P$}
                    \text{discard} $P$ from $S$;
                \EndIf
            \EndFor
        \EndFor
        \\
        $MTS_P \leftarrow S$;\\
        return $MTS_P$;
    \end{algorithmic}
    \label{alg:sky}
\end{algorithm}

Algorithm \ref{alg:sky} takes a list of providers and their corresponding QoS advertisements as input. Initially, we consider each provider is a part of the skyline. Therefore, we create a list $S$ where we assigned each provider. Next, for each provider ($P$) in the list $S$, we compare the provider with all other providers ($P_i$) in $S$ to test whether a provider is dominated by any other provider using the MTS-based dominance relation. If a provider ($P$) is found to be dominated by any other provider ($P_i$) in the list ($S$), we remove $P$ from $S$. Once the algorithm finishes comparing each provider in the list, the remaining providers in $S$ are part of the skyline. $S$ is assigned into $MTS_P$ which represent the final IaaS skyline and returned.

\section{Cooperative Long-term QoS Prediction}
\label{sec:collaborative}
The experiences of trial users may not be directly applicable to predict a provider's long-term performance for a new consumer's workloads. The main reason is that each user may perform a trial with different types of workloads according to its QoS requirements and have different experiences. It is highly unlikely that the new consumer's long-term workloads can be matched exactly with the trial workloads of past trial users at each period. However, the new consumer's workloads may have similarities with the trial user workloads. We utilize the experience of the users having similar trial workloads. 

Collaborative filtering methods are well-known for QoS prediction from similar user experiences  \cite{zheng2012qos}. Traditional collaborative filtering methods for the cloud QoS prediction measure similarities between the users based on their QoS experience. However, they do not consider user workloads during the service invocation. We leverage a traditional collaborative filtering method for the long-term QoS prediction where similarities between a consumer and past trial users are defined based on their workloads. 

Let us assume that the total provisioning period has $S$ seasonal periods. The performance of a provider does not vary significantly within a period. The performance of the provider may vary considerably between the seasonal periods. The total service provisioning time $ T=S \times M \times Y$ where $S$ is the number of seasons, $M$ is the size of each season, and $Y$ is the number of seasonal years. For example, a consumer may need service for one year. The one year can be divided into three seasons where the size of each season is four months. Our target is to predict the QoS performance of a provider for each season for the consumer's workloads at that period. We assume that we have a history of past trial users at each season. The effects of day, night, and week are included within each season. We develop a cooperative long-term QoS prediction \textbf{(CLQP)} approach to predict the QoS performance for each season. The proposed approach has two steps finding similar users and measuring the QoS performance based on similar users' experiences.

\vspace{-2mm}
\subsection{Finding Similar Trial Users}
\label{sec:similar}
A similarity measure technique is required to find similar users based on the consumer's workloads. Pearson Correlation Coefficient (PCC) is an effective similarity measurement technique that has been widely used in collaborative filtering for finding \textit{similar users} \cite{zheng2012qos}. It can achieve high accuracy and can be implemented easily. In a recommender system, similarity between two users is typically computed based on useres' preferences, observed QoS performance, and Location. In our case, we need to find similar users based on the similarity between the workload of two users. For a given interval $\Delta t$, the PCC similarity is between two set of workloads $w$ and $w'$ is defined in Equation \ref{eqn:pearson}: 

\small
\vspace{-4mm}
\begin{equation}
    Sim(w,w')^{\Delta t} = \frac{\sum_{t=j}^k (w'_t - \Bar{w'}) (w_t - \Bar{w})}{\sqrt{\sum_{t=j}^k (w'_t - \Bar{w'})^2} \sqrt{\sum_{t=j}^k (w_t - \Bar{w})^2}}
    \label{eqn:pearson}
\end{equation}

\normalsize

In the Equation \ref{eqn:pearson}, $\Bar{w}$ and $\Bar{w'}$ denotes the average value of $w$ and $w'$ in $\Delta t$. The value of workload $w$ and $w'$ at the timestamp $t$ is denoted by $w_t$ and $w'_t$ respectively.

The PCC may not be always applicable for time series data. Whenever a workload time series segment has a steady workload (i.e., variance = 0), the PCC fails to compute the similarity. We utilize the root mean square error distance to measure the similarity of the consumer workloads using the following Equation \ref{eqn:rmse}:

\small
\vspace{-4mm}
\begin{equation}
\text{Sim' }(w,w') = \sqrt[]{\frac{1}{n}{\sum_{{t=1..n}}  (w_t-w'_t)^2}} 
\label{eqn:rmse}
\end{equation}

\normalsize

Selecting similar users is an important step as selecting dissimilar users may lead to poor QoS prediction. We incorporate an extended Top-K neighbor selection technique to select similar users \cite{tang2016collaborative}. The extended Top-K considers that some users may have a limited number of similar consumers. Traditional Top-K algorithms may select consumers that are not similar to the new consumer. We use the following Equation \ref{eqn:topK} to select Top-K similar consumers:

\small
\vspace{-2mm}
\begin{equation}
   S(c) = \{c_k|c_k \in Top(c); Sim(c_k,c)>0\}
    \label{eqn:topK}
\end{equation}

\normalsize

$Top(c)$ finds the set of similar users $c_k$ to the consumer $c$ and $S(c)$ is the set of similar users who have similarity with the new consumers more than zero.

\subsection{Measuring QoS Performance}

Let us assume that the proposed framework finds $k$ number of users who have trial workloads similar to the new consumer at a given seasonal period. We denote the number of timestamps at a particular season is $n$. For each timestamp $t$, the following to predict the QoS values:

\small
\begin{equation}
    E_{t=1..n}(Q_c) =  \frac{\sum_{i=1}^k Q_{c_k}^{t} }{k} ; t \in \Delta t
    %r_{q'_i,q_i} = \frac{\sigma_{q'_i,q_i}}{\sigma_{q'_i} \sigma_{q_i}}
    \label{eqn:av}
\end{equation}

\normalsize

where $k$ is the number of the similar users, $Q_{c_k}^{t}$ is the observed value of a QoS parameter by a user $c_k$ at time $t$. $E_{t=1..n}(Q_c)$ measures the expected QoS in the trial period $\Delta t$ based on average observed performance by similar users. Equation \ref{eqn:av} considers all users equally without considering the degree of similarity of each user. For instance, if we select five user, a particular user may have the highest similarity. Considering the most similar users will not provide good prediction accuracy. Hence, we consider the \textit{degree of similarity} during the computation of QoS performance for a consumer. The following is used to measure the QoS performance:

\small
\vspace{-3mm}
\begin{equation}
    E'_{t=1..n}(Q_c) =  \frac{\sum_{i=1}^k (Sim'(c,c_k)^{\Delta t}  \times Q_{c_k}^{t}) }{\sum_{i=1}^k Sim'(c,c_k)} ; t \in \Delta t
    %r_{q'_i,q_i} = \frac{\sigma_{q'_i,q_i}}{\sigma_{q'_i} \sigma_{q_i}}
    \label{eqn:predict}
\end{equation}

\normalsize

where $Sim'(c,c_k)^{\Delta t}$ is the similarity between the consumer $c$ and the trial user $c_k$ and $ Q_{c_k}^{t}$ is the observed QoS performance by $c_k$. Equation \ref{eqn:predict} measures the QoS performance using the weighted average of past trial experiences.

% The summation operator ($\sum$) in Eq. \ref{eqn:predict} depends the type of QoS parameter. We use the following QoS aggregation rules introduced in \cite{mistry2016qualitative} to implement the summation operation:   
% \begin{align*}
%       \text{Rule of Summation: Q} = \sum_{i=1}^n Q_i;Q_i \in \{C, M, RP, NB, P \} \\
%     \text{Rule of Minimization: Q} = min_{i=1}^n Q_i;Q_i \in \{A, TP \}
% \end{align*}

% where $C$, $M$, $RP$, $NB$, $P$, $A$, and $TP$ denotes CPU, memory, response time, network bandwidth, price, availability, and throughput respectively. We do not need to divide by $k$ when we apply the minimization rule. The motivation behind using the minimization rule is to predict the minimum QoS performance.
\label{sec:pred}

\subsection{Performing Trial using Workload Replay}

The most straight forward approach to test a provider's performance is to use the consumer's long-term workloads directly to perform the trial. This approach is called \textit{workload replay}. The advantage of this approach is that it performs the most ``real'' test of the providers as the workload contains all the real-world complexities. 

A key challenge of testing is the short length of free trials. A consumer typically requires to test long-term workloads where the trial periods are short. For instance, a consumer may need a VM for one year and the length of the trial period is only one month. The consumer is unable to test its one-year workloads in one month. Let us assume that the trial period is offered in the month of ``June''. The result of performing the trial using the workload of ``January'' may be different from performing the trial using the workload of ``June''. If the consumer performs the trial with the workload of ``June'', then the performance of the provider may remain unknown for the workload of ``January" unless the workloads of both months are the same. A major concern while performing the trial is the effect of the \textit{temporal effect} on the provider's performance. A provider's performance may vary when a consumer performs the trial with the same workload in the ``January'' and ``June''. 

We utilize the consumer's workload \textit{directly} to perform the trial. Trial workloads are selected based on the month of the trial. For example, if the trial is performed in the month of ``June', we apply the consumer's workload of ``June''. When a consumer requires service for multiple years, we take the average workload of multiple years for the trial period. The performance of the workloads for other months is inferred from the experience of past trial users as described in the previous subsections (\ref{sec:similar} and \ref{sec:pred}).  

%We devise two methods for trial workload generation called workload replay and workload compression. The consumer however may not be able to test the total workload in a trial period as the trial period is typically significantly smaller than the long-term service provisioning periods. The consumer requires to apply a smaller portion of the workload in the trial periods.

\subsection{Measuring Confidence}

The predicted QoS performance for the consumer's long-term workloads may not provide an accurate result as it is based on historical information. The current performance of the providers may change. Providers may upgrade their infrastructure or change their QoS management policy. Therefore, the consumer needs to measure the confidence of the trial for the long-term selection.

The intuition behind confidence measuring is that a consumer may have more confidence for the selection if its trial experience matches the experience of other similar consumers. However, each consumer performs trials with different workloads. Therefore, when measuring the confidence, a consumer's experience needs to be compared with other consumers who have similar trial workloads. The confidence of the prediction is measured between a consumer's observed performance in the trial and the predicted performance for the trial period. The predicted performance is computed using the experience of similar consumers. The observed and the predicted performance are both represented as time series. Therefore, we measure the similarity between observed and predicted performance to compute the confidence of the trial experience. Here, we measure the similarity based on \textit{euclidean distance} between the two time series. Euclidean distance is more useful when we want to scale the absolute distance. The confidence of the QoS prediction ($C^{\Delta t}$) computed by measuring the similarity between the trial experience ($E$) and the predicted QoS performance ($Q$) using the following equation:

\small

\begin{equation}
    C^{\Delta t} = \sqrt{\frac{\sum_{t=j}^k (e_t - q_t)^2}{\Delta t}} ; e_t \in E, q_t \in Q \label{eqn:trend}
\end{equation}

\normalsize

where $\Delta t$ is the length of the trial period, $E$ is the observed QoS performance, $Q$ is the predicted QoS performance in the trial period. $e_t$ and $q_t$ is the obsereved and predicted performance at timestamp $t$. The Equation \ref{eqn:trend} measures similarity in terms of the euclidean distance of the predicted and observed QoS performance. The proposed long-term selection framework selects a threshold $C^{thres}$ for the confidence variable to filter providers based on the confidence. If the prediction of a provider has lower confidence than the threshold, the provider is discarded.

\section{Long-term QoS Prediction without Historical Information (LQP-short)}

The workload replay based trial is not suitable for performing a trial in the absence of similar trial users. The consumer needs to measure the IaaS performance for its entire workloads. We propose a trial workload generation model using time series compression techniques.

\subsection{Trial Workload Generation}

The performance of a cloud provider often depends on the characteristics of the workloads it serves. For instance, the performance of Amazon AWS shows less performance variability for compute-intensive tasks according to a report made by Cloud-Spectator\footnote{https://cloudspectator.com/cloud-performance-reports/}. The performance of Amazon AWS varies considerably for block storage operations. The distribution of the workloads may have a considerable impact on the performance. A provider's performance may show better performance when the workloads are evenly distributed. If the workloads come in burst or have heavy-tailed distribution, the performance may degrade \cite{al2018performance}.

Let us consider that the long-term workloads of the consumer consist $n$ data points, i.e., $t_1, t_2, t_3,...,t_n$ over $T$ period. The trial period can be divided into $k$ data points over $T_r$ period where $T_r << T$. If $n \leq k$, then each data point in the workload can be tested in the trial period. If $n > k$, then there are more workload data points than can be tested and compression is required. In such a case, the compression may result in some loss of data points. The amount of loss depends on \textit{the size of $k$}, \textit{the shape of the workload time series}, and \textit{the compression method}.

We define $<W,M,T_r,k,Q_C>$ as a trial workload generation model where $W$ denotes the long-term workload time series of a consumer, $M$ is the compression method for the workload time series, $T_r$ is the trial period offered by a provider, $k$ is the number of points in $T_r$, and $Q_C$ is the list of QoS parameters that need to be measured in the trial. The goal of this model to compress the workloads $W$ using method $M$ into $k$ points to fit in $T_r$.

There exist a number of techniques for time series compression \cite{burtini2013time}. We apply three techniques namely, Piecewise Uniform Selection(PUS), Piecewise Aggregate Approximation (PAA), and Random Selection (RS). 

\begin{enumerate}[itemsep=0ex, leftmargin=2ex]
% \item \textit{Perpetually important points (PIPs):}  Perpetually important points (PIPs) is a well-known algorithm proposed in \cite{fu2005adaptive}. PIP algorithm determines most visually important points for a given time series. It takes a time series ${t_1,t_2,....,t_n}$ as an input and draws a line segment from $t_1$ to $t_n$. It finds the farthest point $t_i$ from $\overline{t_1t_n}$. If the distance is more than a predefined error $\epsilon$ than, the algorithm recursively approximates the subintervals $t_1,t_2,...,t_i$ and $t_i,t_{i+1}...,t_n$. The number of point to be generated by the algorithm determined by the $\epsilon$. 

\item \textit{Piecewise Uniform Selection (PUS):} The original workload time series is divided into $k$ intervals \cite{burtini2013time}. Starting with the first workload, we select a workload point after each $\lceil n/k \rceil$ interval. The shape of the original workload time series may remain the same if the workload does not vary significantly over time. 

\item \textit{Piecewise Aggregate Approximation (PAA):} The PPA method reduces the number of data points by taking average values in each interval $I_i$. If $t_i$ represents a timestamp in the workload time series and $n$ is the total number of points in the time series, then the value of the time series in $I_j$ is calculated using the following equation: 

\small

\begin{equation}
I_j = \frac{1}{x} \sum_{i=(j-1)*x+1}^{i*x}t_i  \text{ for } j = 1...\lceil n/x  \rceil
\label{eqn:compress}
\end{equation}

\normalsize

The size of the original time series can be reduced by any factor by changing the value of $x$. For a given $n$ data points in the workloads and $k$ segments in the trial period, we define the minimum $x=\lceil n/k \rceil$. 

\item \textit{Random Selection (RS):} The trial workloads are generated by selecting $k$ number of workloads randomly from the original workloads. The random numbers are generated based on a uniform distribution in the interval (0,1).

\end{enumerate}

\subsection{Long-term QoS Prediction}

The proposed framework monitors the required QoS parameters of the consumer ($Q_C$)  during the trial period. The proposed framework expands the short-term QoS performances into the long-term performance using \textit{time series interpolation} and \textit{extrapolation}. 

For a given set of workloads $W=\{w_1,...,w_n\}$ and $k$ number of points in the trial period $T_r$, the compression method $M$ generates a set workload points $W'=\{w'_1,w'_2,..,w'_k\}$ for the trial. For each of these workloads, the corresponding QoS value is monitored in the trial. There are $k$ number of values for each QoS parameter $q_{ci} \in Q_C$. There exists a one-to-one mapping between each $w_i$ to $x_j \in q_{ci}$. The aim is to generate $n$ number of QoS values from $k$ number of QoS values of each $q_{ci} \in Q_C$.

The proposed framework leverages the knowledge of compression to expand the QoS values. The workload time series is divided into $\lceil n/k \rceil=x$ intervals during compression. The aim is to find $n-k$ points to determine the long-term performance. We apply the interpolation and extrapolation method to generate $n-k$ points. Interpolation is a method is a well-known technique in the field of numerical analysis to construct new intermediate points between two points in a time series \cite{wiener1949extrapolation}. First, $n-k$ workload points are mapped into the different points of the time series. The mapped workload points are interpolated and extrapolated to generate intermediate workload points.

There are various algorithms to perform interpolation and extrapolation such as piece-wise constant, linear, nearest neighbor, and polynomial interpolation \cite{wiener1949extrapolation}. We select a commonly used interpolation technique, i.e., linear interpolation. If two consecutive workload points are denoted by $(t_1,w_1)$ and $(t_2,w_2)$, then an intermediate point is calculated by the following equation:

\small
\vspace{-2mm}
\begin{equation}
w=w_1+(t-t_1) \frac{w_2-w_1}{t_2-t_1}
\end{equation}

\normalsize

\section{QoS-aware Long-term IaaS Provider Selection (QLIS)}
The proposed framework selects the providers based on the confidence of the prediction when similar trial users are available. First, a subset of candidate providers is selected based on the confidence threshold ($C^{thres}$). This step is not performed when similar trial users are not available. Next, the proposed framework ranks the providers based on the distances of the predicted QoS performances and the consumer's QoS expectations. The predicted QoS performance and the consumer's QoS expectations both are represented using TSGs. Measuring the similarity between two TSGs can be costly when the number of QoS parameters is large. The cost of the pairwise comparison of the time series can be reduced if we can lower the dimension of the TSGs.

%\cite{mackiewicz1993principal}
Principal Component Analysis (PCA) is an efficient approach to reduce the dimension of a TSG by transforming the original TSG into lower dimensional vector space \cite{mackiewicz1993principal}. The PCA transformation is represented $D'=D \times T$ where $D$ is the original data, $T$ is the transformation vector, and $D'$ is the transformed data matrix into the new vector space. Each data vector of the original space is represented by a column of $D$. Each row of $T$ refers to a transformation vector. Each column of $D'$ represents a principal component of the original vector space. 

Let us consider that the set of discovered QoS is represented as $TSG$ which is a $m \times n$ dimensional vector where $m$ is the number of timestamps and $n$ is the number of QoS parameters in the TSG. A vector $Q_t=(q_1^t,q_2^t,....q_n^t)$ is created for each timestamps. Each dimension of the vector refers to a QoS time series in $TSG'$. We apply PCA transformation on the vector to identify the first $n'$ QoS time series where $n'<n$ to get a new TSG $TSG'$. This $n'$ contains the principal components of the discovered QoS performances. The transformation is computed as follows:

\small

\begin{equation}
     D'_t= D_t \times T = (D_t \times T_1, D_t \times T_2, .. D_t \times T_n)
\end{equation}

\normalsize
    
where $D_t$ is the original data, $D'_t$ is the new data, and $T_i$ is a n dimensional transformation vector. The values of $D'_t$ for each timestamp is the first $n'$ principal components of $D_t$. Given a $n'$ dimensional transformation matrix $T'=(T_1, T_2, .., T_n')$, the transformed TSG is as follows:

\small

\begin{equation}
    TSG' = D \times T' = (D_1 \times T', D_2 \times T',...., D_m \times T')
\end{equation}

\normalsize

where $m$ represents the number of timestamps in each time series. Given the QoS requirements of the consumer $TSG1$, the transformed QoS performances $TSG2$ of a provider, and the number of the time series is $n$, the distance is calculated by the following Equation \ref{eqn:distance}:

% Equation \ref{eqn:mea} one of the technique to compare the similarity between two time series by calculating the absolute mean difference.

% Measuring the similarity over multiple QoS time series is challenging. If the number of timestamps is large, the pairwise comparison may be very costly. 
%PCA is a commonly used time series dimensional reduction technique.  

% Once the performances of the providers are predicted, we rank the providers based on their discovered performances and the consumer requirements. There are different techniques to measure the similarity between a set of time series. Equation \ref{eqn:loss} can be used to compare the similarity between two time series by calculating the absolute mean difference. Measuring the similarity over multiple QoS time series is challenging. If the number of timestamps is large, the pairwise comparison may be very costly. We use Principal Component Analysis (PCA) \cite{mackiewicz1993principal} to reduce the number of time stamps of a time series. PCA is a commonly used time series dimensional reduction technique. 

\small

\begin{equation}
\text{distance }(TSG1,TSG2) =  \sum_{i=1}^n \text{RMSE} (Q_i^1,Q_i^2)
\label{eqn:distance}
\end{equation}

\normalsize
The proposed framework ranks the providers based on normalize RMSE distances of their QoS performances with the consumer requirements.

\section{Experiments and Results}

We conduct a set of experiments to evaluate the proposed framework. First, we investigate the dominant QoS attribute-based filtering process and compare it with the MTS-based temporal skyline proposed in \cite{wang2013dominant}. Next, we evaluate the effect of the proposed trial workload generation approaches (i.e., PAA, PUS, RS) on the long-term performance prediction. We then assess the performance of the proposed CLQP approach and the LQP-short approach. Finally, we evaluate the ranking accuracy of the proposed framework and compare it with the traditional trial-based ranking approaches \cite{li2010cloudcmp, wang2018testing}.

\subsection{Experiment Requirements}

\begin{figure}
 
 \centering
 
 \includegraphics[width=0.3\textwidth]{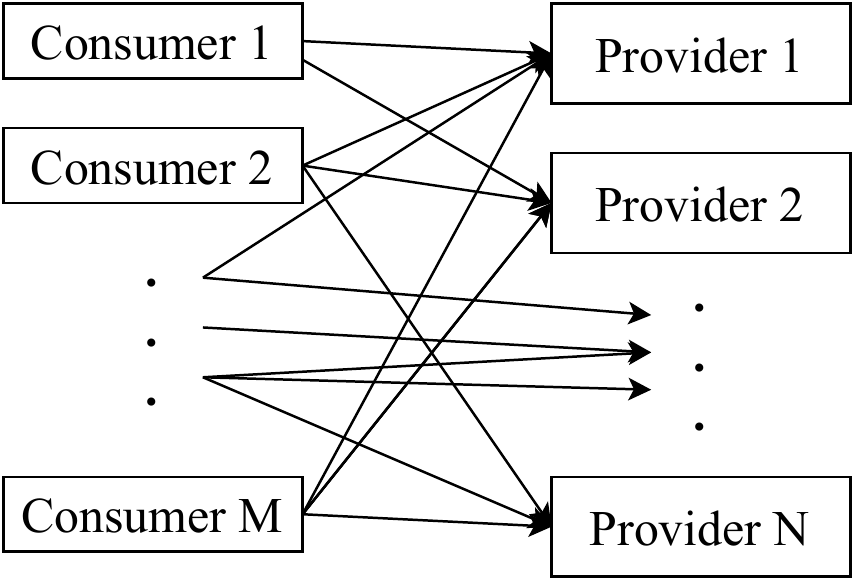}

    \caption{Experiment Settings}
    \vspace{-5mm}
    \label{fig:exp_scene}
\end{figure}

To evaluate the proposed framework, we require an environment where a set of IaaS providers advertises their long-term performance, a set of consumers who performs trials on different providers based on their workloads over different periods. Figure \ref{fig:exp_scene} depicts such an environment. Each consumer runs the same workload on different providers to decide which provider is the best for their workloads. The proposed framework would help a new consumer to select the closest match provider according to its QoS requirements. First, the proposed framework filters providers based on their advertisements. Next, it generates trial workloads based on the consumer's long-term workload to discover providers' performance in the trial period. Next, the framework utilizes the experience of past trial consumers to discover providers' performance on different periods outside of the trial period. Finally, it ranks the provider based on the discovered performance and the consumer's QoS requirements. Therefore, we require advertisements for a set of IaaS providers, trial workloads of a set of consumers, QoS performance of providers on different periods for the consumers' trial workloads. In addition, we require a consumer's long-term workloads to perform the trials and the long-term performance of providers for the consumer's long-term workloads as ground truth to evaluate the proposed framework.

\subsection{Experiment Design and Data Preparation}

Finding real-world datasets for a long period that meet our experiment requirements is challenging. In particular, we require IaaS cloud workload traces and their corresponding performance datasets for a set of providers for a long-term period. To the best of our knowledge, there is no existing publicly available long-term workload-performance datasets of IaaS cloud providers. Therefore, we leverage existing short-term available datasets to synthesize datasets for our experiments. In particular, we use Eucalyptus IaaS Cloud workloads \cite{wolski2017qpred}, and SPEC 2016 performance benchmark results \cite{baset2017spec} to create experimental datasets. In the following subsections, we provide a brief description of the datasets that we used for the experiments and the data generation process to conduct the experiments. \textit{We have made our dataset and source code publicly available to make this experiment reproducible}\footnote{\url{https://github.com/sm-fattah/IaaS_cloud_experiment}}.

\subsubsection{IaaS Advertisements Generation}

IaaS advertisements in the real-world mostly contain short-term price and availability information. Therefore, we randomly create 60 IaaS advertisements which contains price, availability, throughput, and response time to understand the effect of a different number of dominant QoS parameters in the filtering process.

\subsubsection{Workload Datasets Preparation}

We utilize the Eucalyptus IaaS workload traces \cite{wolski2017qpred} to represent past consumers' trial workloads and a new consumer's long-term workloads. To the best of our knowledge, Eucalyptus private cloud usage traces are the closest match to validate the proposed approach with real-world datasets. The Eucalyptus traces has been utilized in several recent studies to represent a real-world cloud environment \cite{wolski2017qpred,pucher2016using}. Eucalyptus published workload traces of private cloud services, which is leveraged by several large companies \cite{pucher2015using}. The published datasets are anonymized multi-month traces scraped from the log files of 6 different production systems running Eucalyptus IaaS clouds. The trace contains information about both the IaaS service level (i.e., VM instance requests, request timestamps, instance lifetime, and service usage time periods) and physical level (i.e., number of cloud nodes, number of CPU cores at each node, and occupancy of physical hosts).  We select a trace called ``D6trace'' which contains VM usages history of a large company with 50,000 to 100,000 employees \cite{pucher2015using}. It contains approximately 34 days of workloads for 31 cloud nodes where each node contains 32 cores. We selected 30 cloud nodes to represent the trial workload of past consumers and 1 cloud node to represents a new consumer's long-term workloads. The workload contains 6486 timestamps. To simplify the experiment, we utilized the piece-wise aggregate approximation to create 360 timestamps. We assume that each workload at a timestamp represents average CPU requests of a consumer on that timestamp. The description of the workload dataset is given by Table \ref{tab:data}.

\begin{table}

\caption{Workload Trace Description} \label{tab:data}
\centering
\begin{tabular}{|l|r|}
 \hline
 Parameter & Value \\
 \hline
 Number of cloud nodes & 31 \\
 Number of CPU cores at each node & 32 \\
 Number of timestamps & 6486 \\
 Resource allocation unit & number of CPU cores \\
 Trace attributes & VM start, stop timestamps, \\
                 & VM resource requests, \\
                 & VM id, node id  \\
 \hline
 \end{tabular}
\vspace{-4mm}
\end{table}

\subsubsection{Workload-Performance Dataset Generation}

\begin{figure}[b]
  \vspace{-5mm}
 \centering
 
 \includegraphics[width=0.35\textwidth]{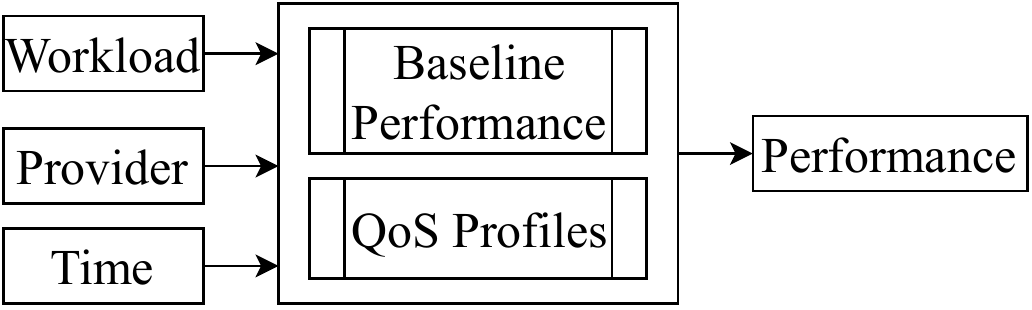}

    \caption{Data Generation Framework}
   
    \label{fig:data_gen}
\end{figure}

Finding performance datasets of a long-term workload trace is very challenging. Therefore, we decide to synthesize the performance dataset for our selected workload traces. We develop a performance data generation framework that generates QoS performance of a set of providers for a given workload for a given period of time. We consider performance data for three QoS attributes based on the available dataset, i.e.,\textit{ CPU throughput (operations/seconds), disk read and insert time (milliseconds)}. Figure \ref{fig:data_gen} shows the performance data generation framework. The framework takes a consumer's workload $W$ for a given period $T$, and a provider's id $P$ as inputs. It generates corresponding QoS performance using two modules a) Baseline performance, and b) QoS profiles of a set of IaaS providers. The baseline performance module maps each unique workload request of the eucalyptus trace to a particular performance value. Therefore, the baseline performance represents the initial performance of a provider for a given workload. The QoS profiles module determines the final performance value of the workload at a given time for the providers based on their QoS profiles. We performed the following steps to generate the baseline performance and QoS profiles:

\noindent \textbf{1) Baseline Performance:} The baseline performance is generated from the benchmark results published by SPEC Cloud IaaS 2016 \cite{baset2017spec}. SPEC Cloud IaaS 2016 benchmark results contain the CPU throughput (op/sec), disk insert and read response time (ms) measurements of private cloud providers. We have collected approximately 1500 performance data observations from the benchmark result. To create a \textit{workload-QoS map}, we map each unique workload request to a unique performance value based on the resource consumption of the requests. A workload with the highest resource consumption is mapped with the lowest QoS performance value. The lowest resource consumption workload is mapped with a high QoS performance value. The baseline QoS performance is utilized to generate long-term performance of IaaS providers with the help of \textit{QoS profiles} of each provider.

\begin{figure}
\centering
\includegraphics[width=.3\textwidth]{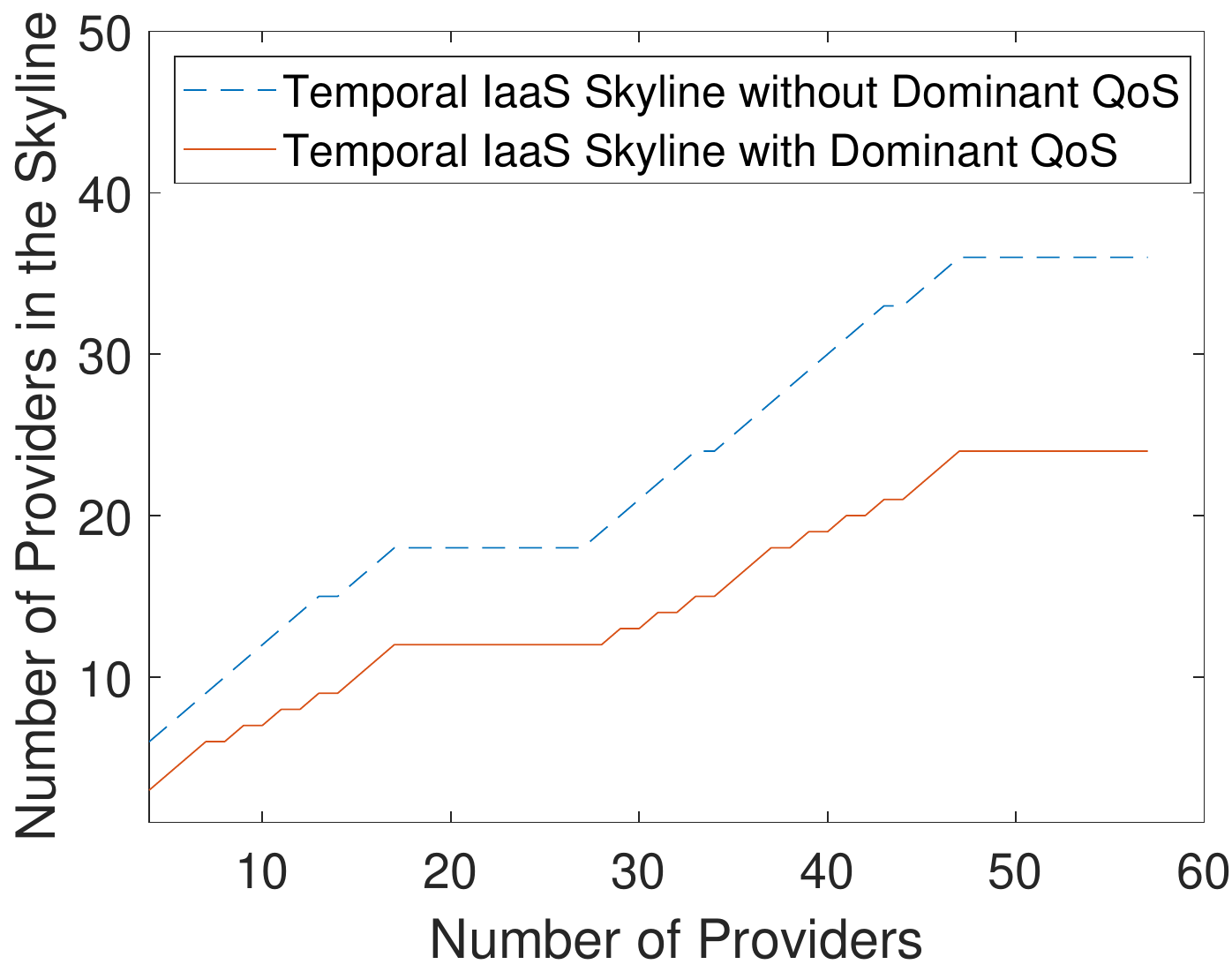}
\caption{Temporal IaaS skyline}
\label{fig:skyproviders}
\vspace{-6mm}
\end{figure}

\noindent \textbf{2) QoS Profiles:} We decide to create QoS profiles of 5 IaaS providers to simulate a real-world environment. A QoS profile determines what would be a provider's performance for a given workload at a certain point of time compared to the baseline performance. For example, if a workload with a low resource consumption is given in January to a provider, it offers 20\% additional throughput compared to the baseline performance. Each QoS profile consists of two maps a) workload map, and b) seasonal map. The workload map determines what would be the expected performance for a given workload compared to the baseline performance. Similarly, the seasonal map determines what should be the performance for a given time compared to the baseline performance. We have created the workload map and the seasonal map randomly. The rules of QoS profiles also add randomness to simulate a real-world provider who may give different performances for the same workload at the same time. The consumer and trial users are unaware of the QoS profile. They may observe the performance variability of the providers. Finally, these QoS profiles of providers are applied to the baseline performance of a given workload to generate the performance dataset of each provider.

We input the trial workloads of past consumers and a new consumer's long-term workloads from the eucalyptus trace into the data generation framework. The framework generates performance data for each provider based on their QoS profiles and the baseline performance. The performance of the long-term workloads is utilized as the \textit{ground truth} for the evaluation.

\subsection{Experiment Result Analysis}

\subsubsection{Effect of Dominant QoS Parameters on IaaS Skyline}

\begin{figure*}[htb]
\vspace{-5mm}
    \centerline{
      \subfloat[]{\includegraphics[width=0.42\textwidth]{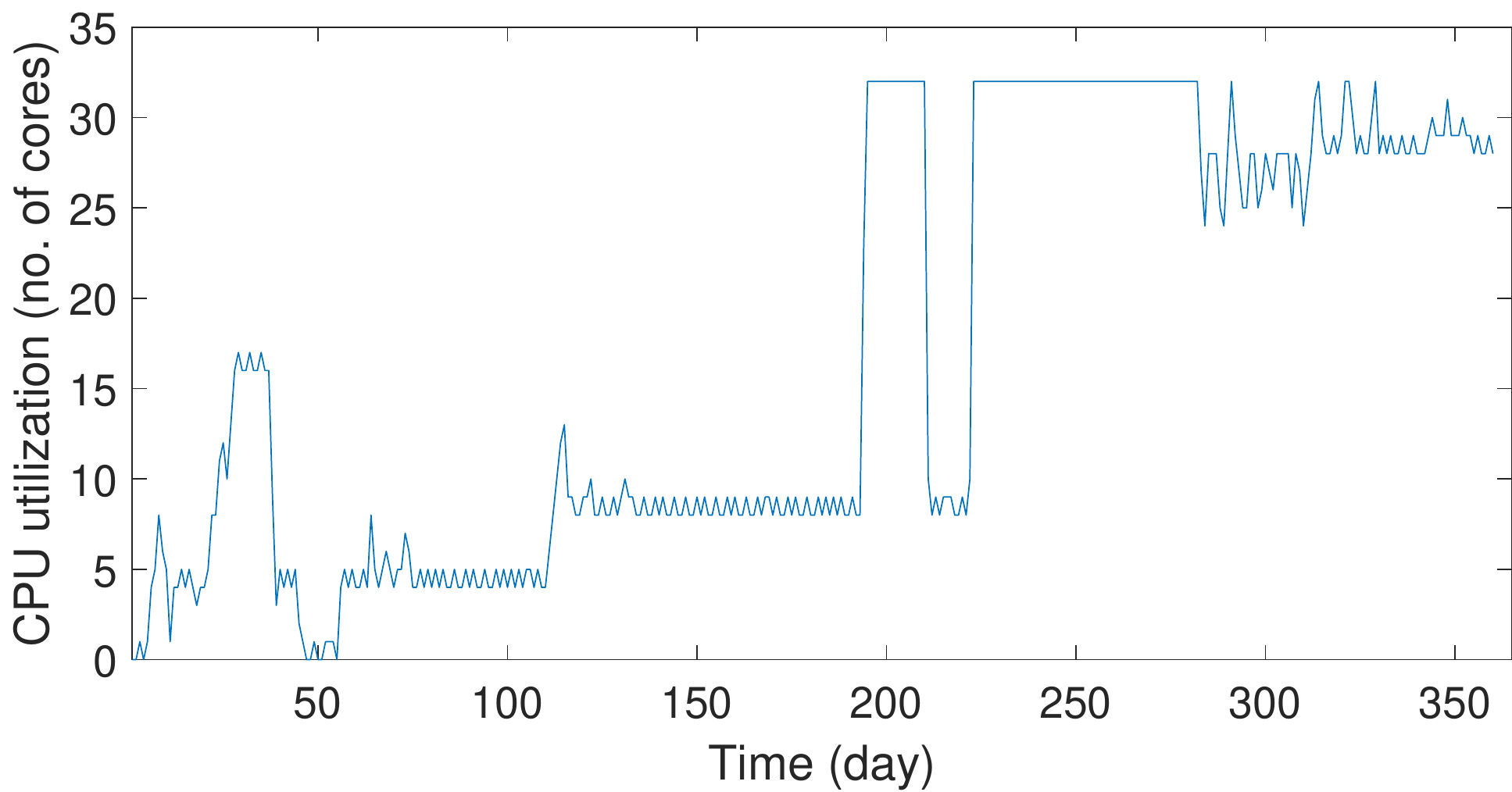}}
      \hfil
      \subfloat[]{\includegraphics[width=.45\textwidth]{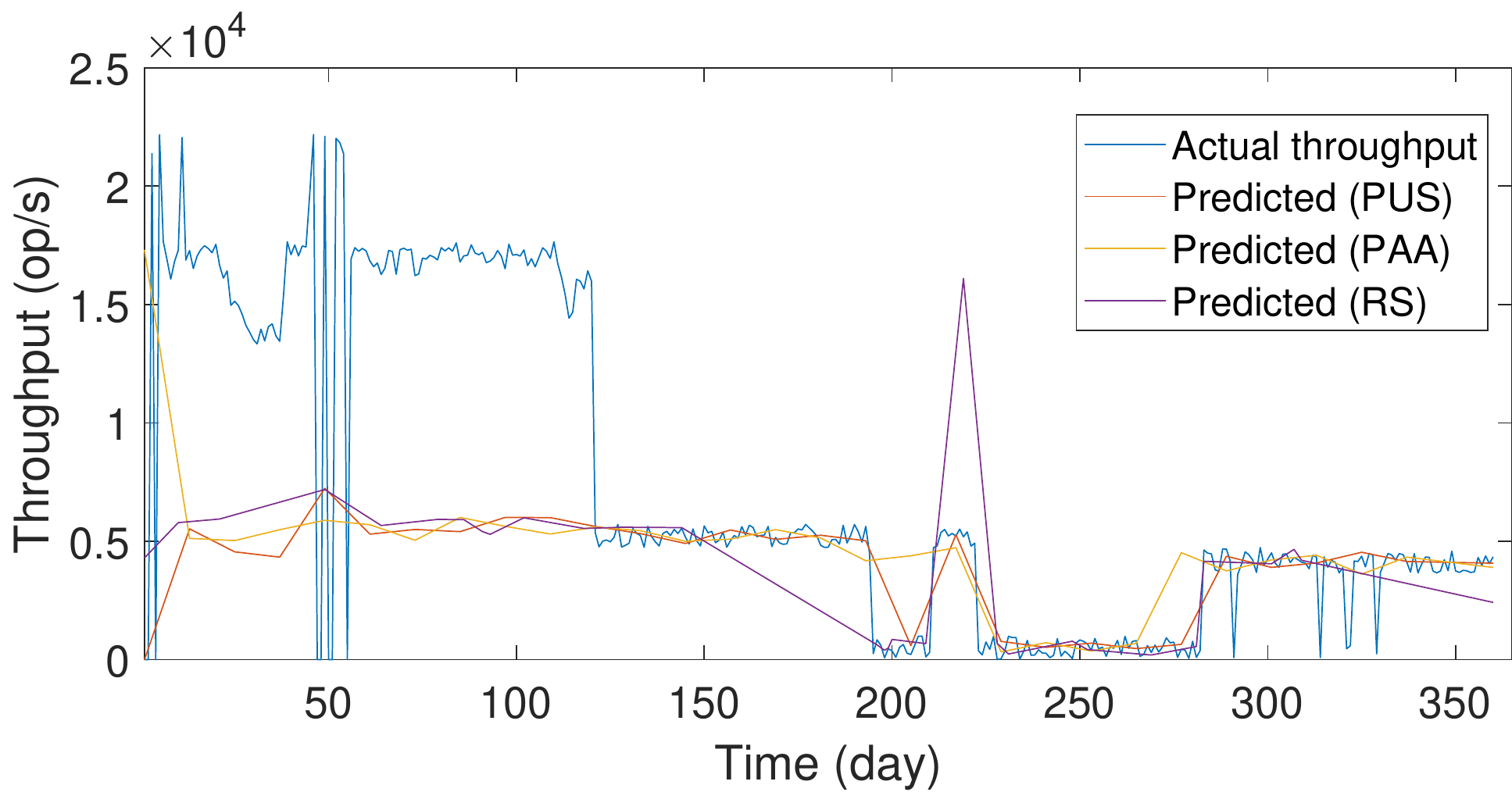}}
      
    }
    \caption{Performance discovery using LQP-short (a) Consumer's long-term workloads (b) Actual and predicted throughput of an IaaS provider}
    \label{fig:pred_without}
    \vspace{-5mm}
\end{figure*}

We apply the filtering on 60 IaaS providers for four QoS parameters. We select the price and availability as dominant QoS parameters. Figure \ref{fig:skyproviders} compares the performance of the temporal IaaS skyline without the dominant QoS parameters and with the dominant QoS parameters. The size of the temporal skyline without Dominant QoS parameters increases considerably with the number of providers. If there are too many IaaS providers in the cloud market, the temporal skyline without the dominant QoS parameter may not be very useful. It only discards 24 providers when the number of providers is 60. The temporal IaaS skyline with the dominant QoS parameters has a considerable impact on the total number of skyline IaaS providers. It always filters more IaaS providers than the temporal IaaS skyline. The main reason is that the size of the IaaS skyline may grow exponentially with the number of QoS parameters. The dominant QoS parameters keep the size of the skyline considerably low. The proposed method successfully reduces the number of IaaS candidates for the trial.

\subsubsection{Evaluation of LQP-short approach}

Figure \ref{fig:pred_without}(a) shows a consumer's workload for 360 timestamps. The average requested number of CPU cores per timestamp is shown in the figure. Three types of trial workloads are generated using three compression techniques (PUS, PAA, and RS). The trial QoS performances are generated for each provider based on their QoS profiles using the data generation framework. The LQP-short mainly utilizes the trial QoS performances to predict each provider's long-term QoS performance.

Figure \ref{fig:pred_without}(b) shows the actual throughput and the predicted throughput of a provider based on the LQP-short and three trial workload generation approaches. The provider exhibits variable performance for the same workload according to the actual performance graph. There are substantial performance fluctuations over different timestamps. The predicted QoS performance exhibits random behavior. The effect of different trial workload generation technique is not much noticeable. The RS based prediction shows more fluctuations than the other two approaches. The QoS predictions seem better when the consumer's workloads are steady (200 to 300 days) as the trial is performed in that period. Apart from the trial period, the LQP-short approach seems to perform poorly. None of the trial workload generation approaches helps to capture the temporal shift in the performance. The main reason for such lower accuracy is because the prediction is performed without utilizing past history. The LQP-short can be considered as a best-effort approach when there are no similar users available.

\begin{figure}

    \centerline{
        \subfloat[]{\includegraphics[width=0.22\textwidth]{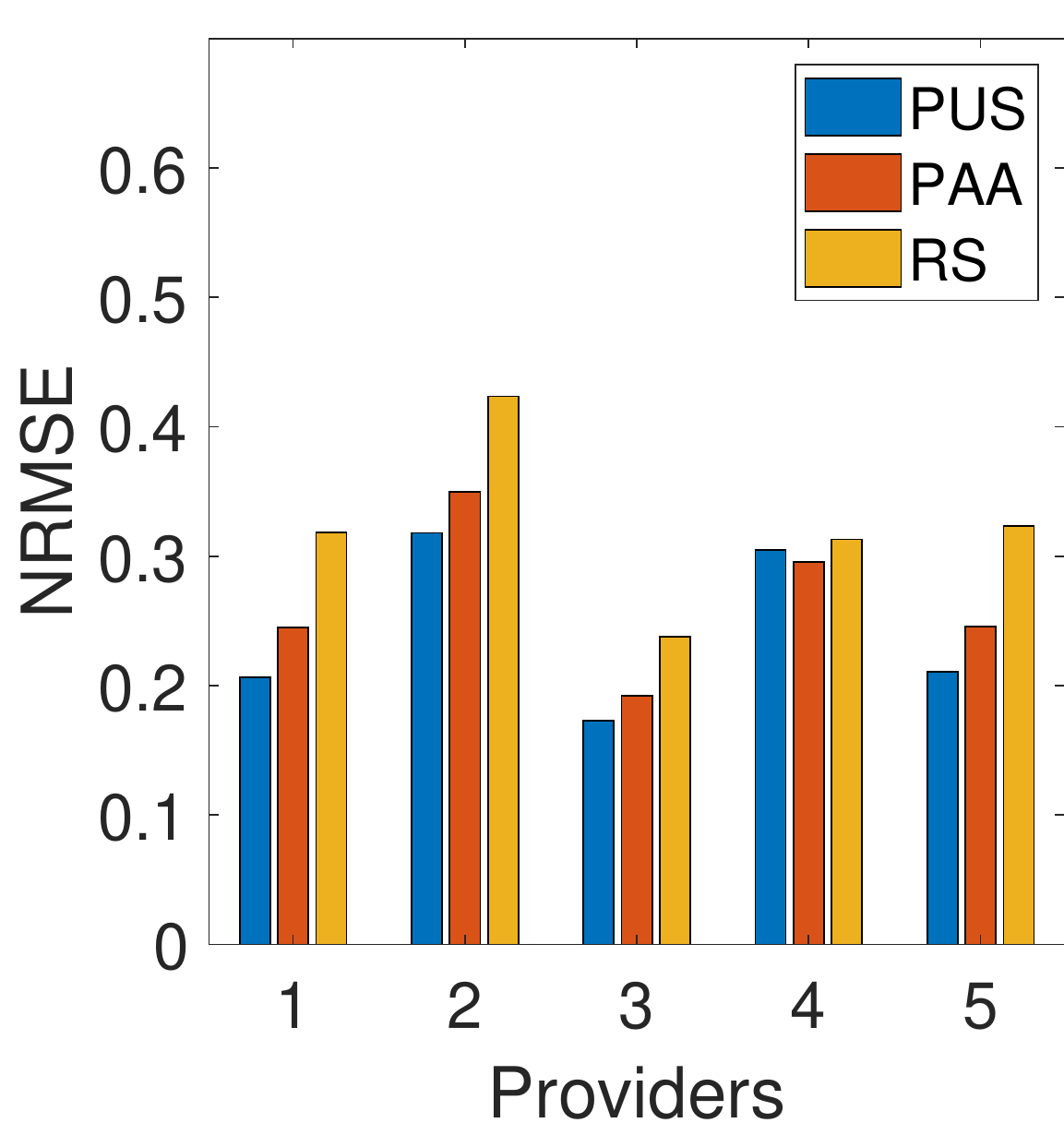}}
      \hfil
      \subfloat[]{\includegraphics[width=0.22\textwidth]{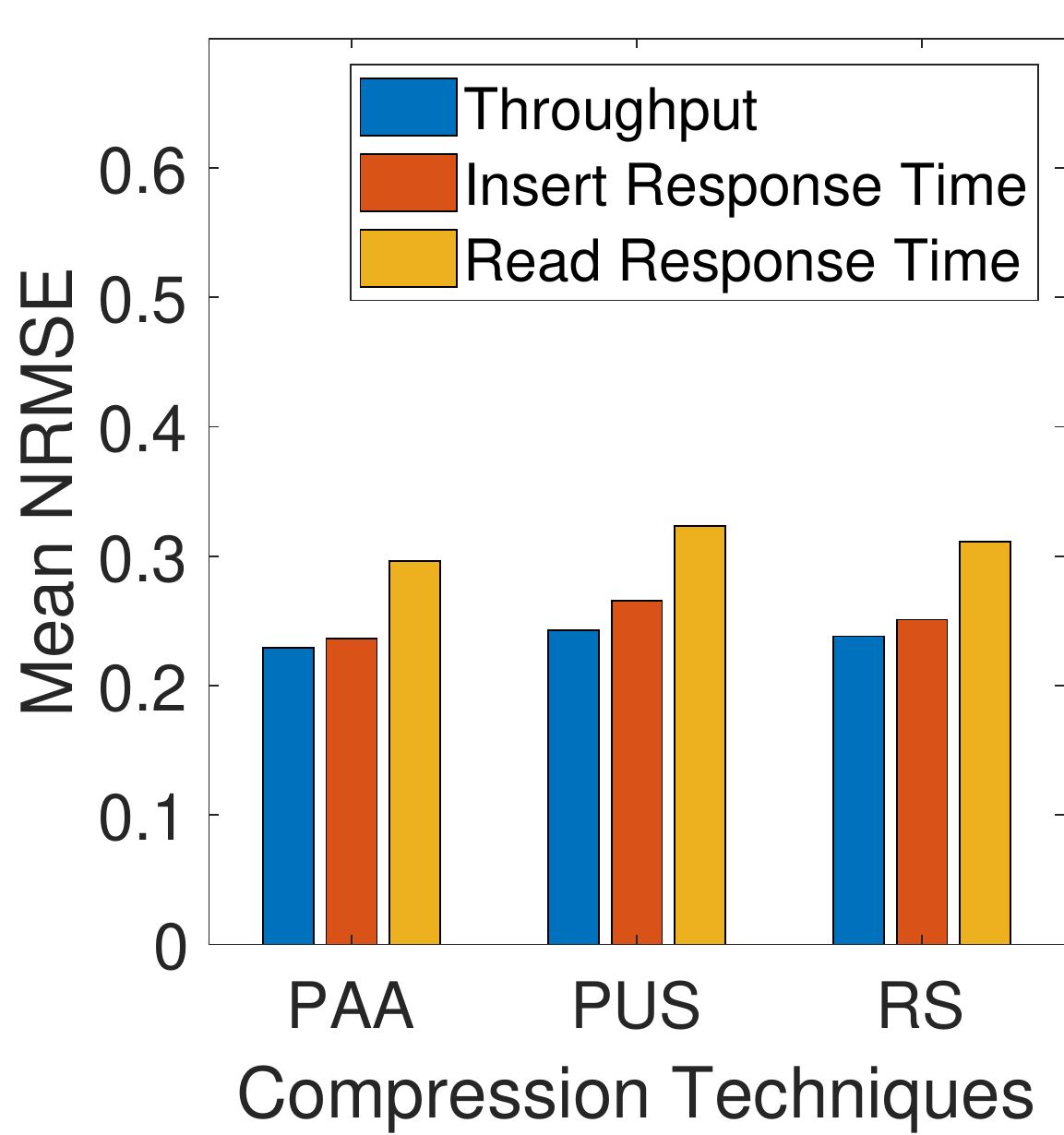}}
    
    }

    \caption{Prediction accuracy (NRMSE distance) (a) Throughput prediction (b) Mean prediction accuracy}
    \label{fig:predErrorTP}
    \vspace{-5mm}
\end{figure}

The prediction accuracy of the LQP-short approach is presented in Figure \ref{fig:predErrorTP}. The throughput prediction accuracy of each provider is shown in Figure \ref{fig:predErrorTP}(a) for each trial approach. The accuracy of the RS method is unpredictable across the providers as the workloads are selected randomly. The RS method has the highest NRMSE distance compared to the other two approaches for each provider. \textit{This confirms that an unplanned short-term trial may lead to poor decision making.} The performance of the PUS approach is consistent with each provider. The PUS method performs better as it retains the shape of the original workloads better than the other two approaches. Hence, choosing the type of workloads is has a considerable impact on trial prediction accuracy. Figure \ref{fig:predErrorTP}(b) shows the mean prediction accuracy across five providers for the QoS parameters throughput, insert response time, and read response time.  The trial workloads generated by the RS method has the maximum NRMSE distance for throughput and read response time. The PUS has the minimum NRMSE distance as it performs well for each QoS parameter.

\subsubsection{Evaluation of CLQP approach}

\begin{figure*}
\vspace{-5mm}
    \centerline{
    
    \subfloat[]{\includegraphics[width=.49\textwidth]{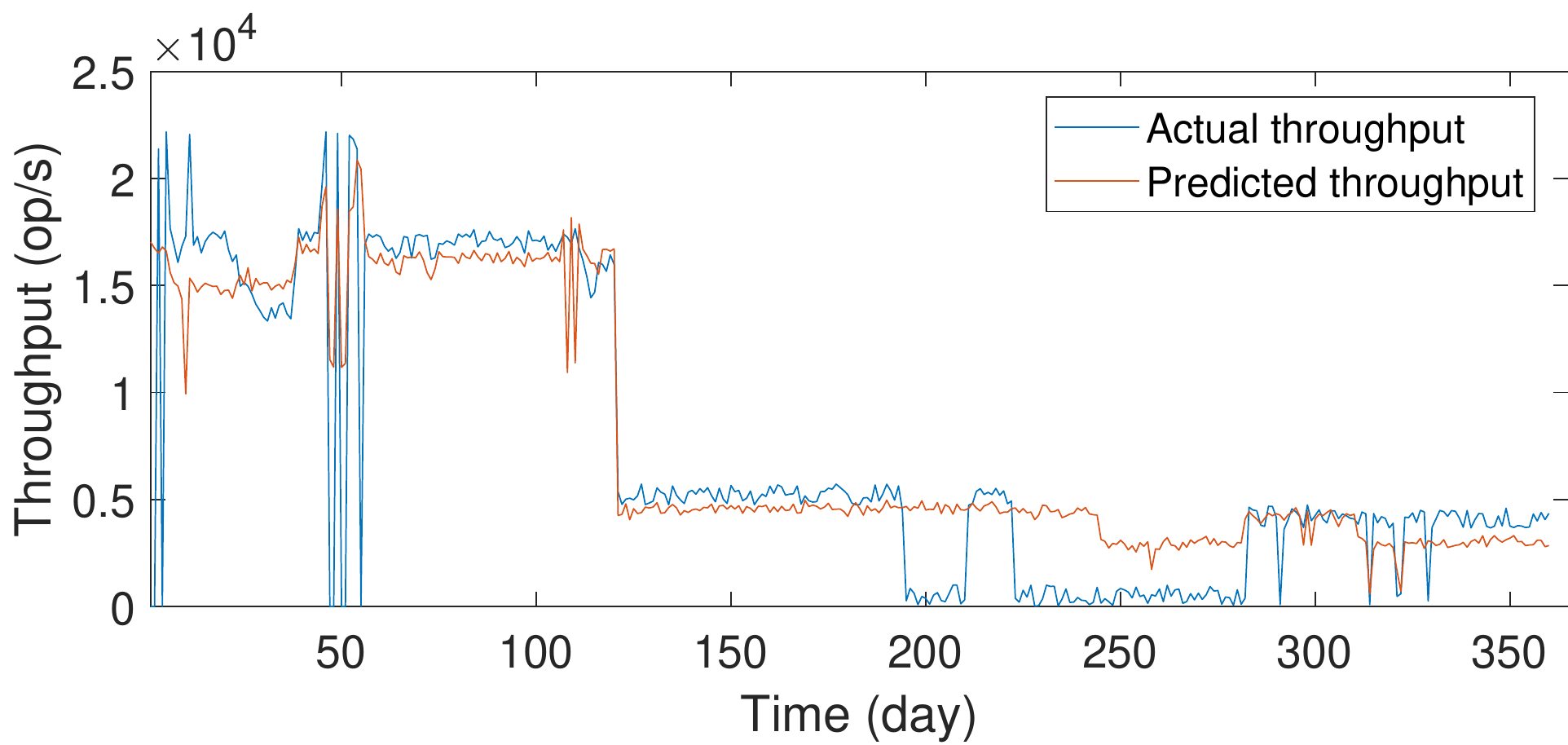}}
    \hfil
    \subfloat[]{\includegraphics[width=.23\textwidth]{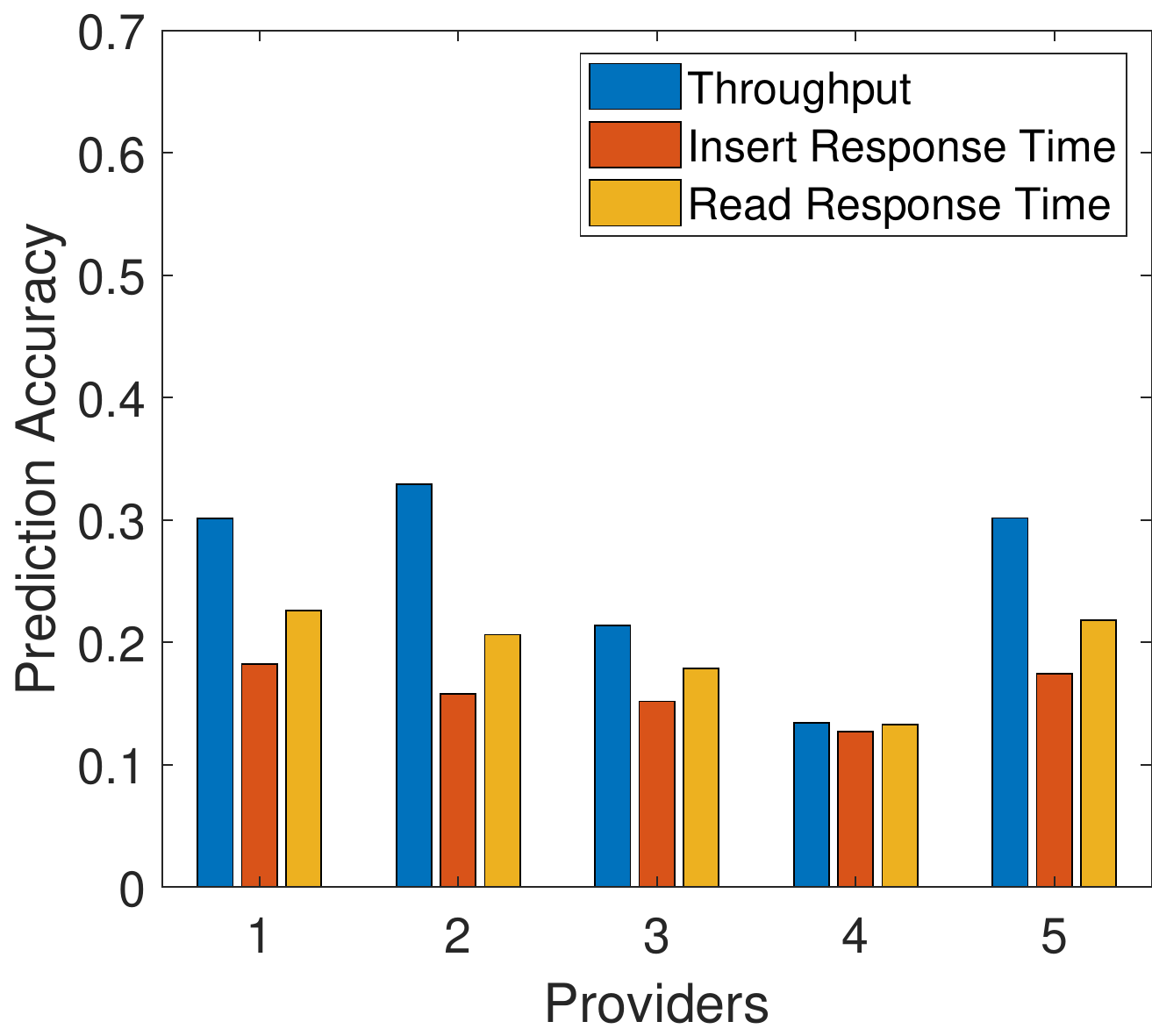}}
    \hfil
    \subfloat[]{\includegraphics[width=0.23\textwidth]{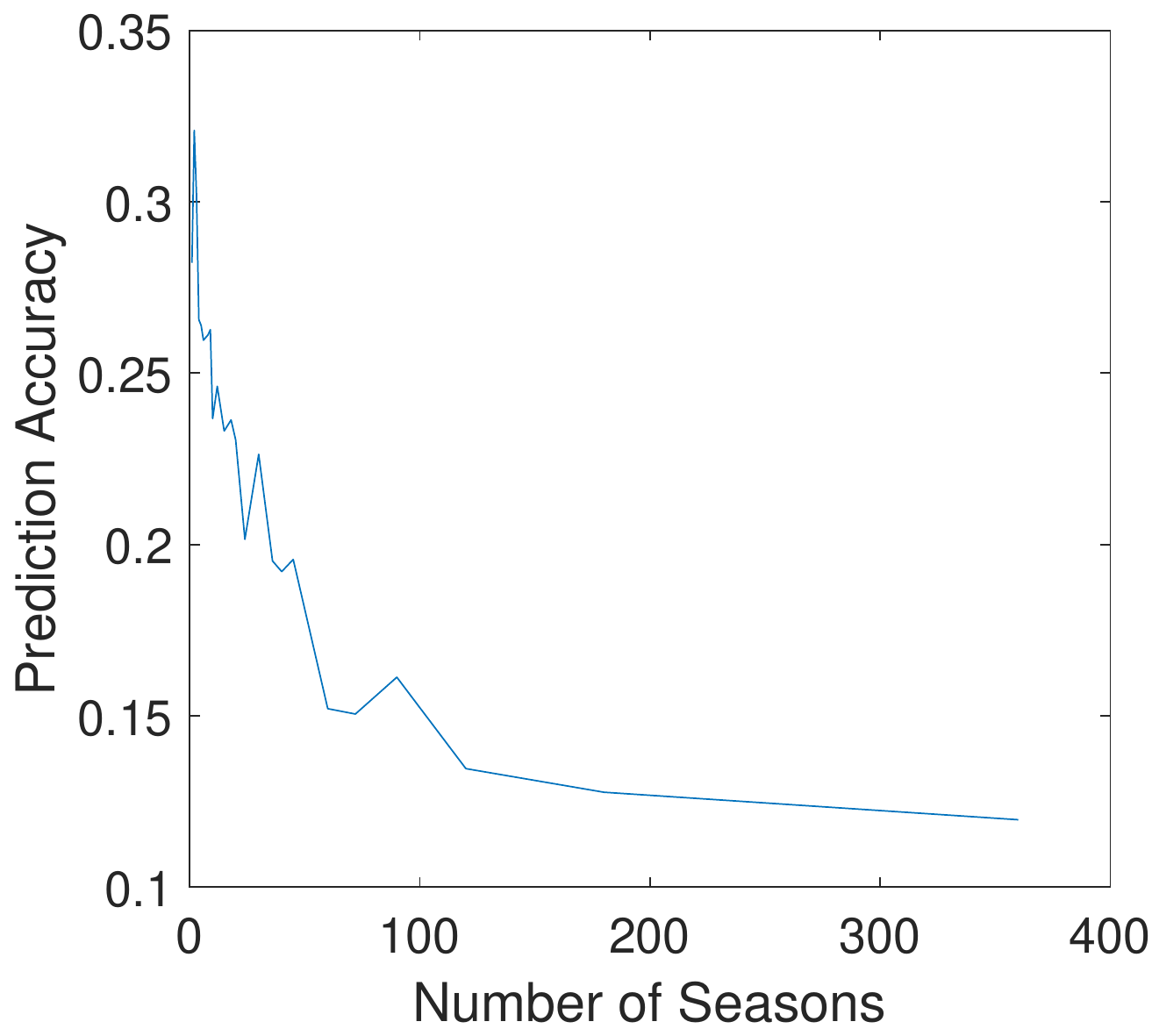}}

    }

  \caption{Performance discovery using CLQP (a) Actual and predicted throughput of a provider (b) Prediction accuracy of the CQLP approach (C) CQLP accuracy for variable temporal segments}
\label{fig:copPred}
    \vspace{-5mm}
\end{figure*}

We divide the total provisioning time into smaller segments to find similar trial users. The provisioning time is divided into 12 temporal segments to perform the cooperative prediction where each segment is 30 days. We run the CLQP using past trial user experiences and the consumer's long-term workloads for each month. Figure \ref{fig:copPred}(a) illustrates the results of the cooperative long-term QoS prediction for an IaaS provider. The accuracy of the prediction improves considerably than the prediction without the history in Figure \ref{fig:pred_without}(a). The prediction effectively captures the temporal and workload performance variability of the provider. The randomness of the performance, i.e., different performance for the same workload at the same time is also captured effectively in the prediction. The accuracy of the predictions is shown in Figure \ref{fig:copPred}(b) in terms of NRMSE distance. The throughput of each provider has a lower prediction accuracy as the NRMSE distance is high. The read response time for each provider has the highest prediction accuracy for each provider. It implies the effect of the performance variability may depend on the type of QoS attribute.

Our intuition is that the number of temporal segments may have a considerable impact on accuracy. When each temporal segment has a smaller length, the probability of getting a similar trial user increases. We conduct the experiment by changing the size of the temporal segments for the cooperative prediction. Figure \ref{fig:copPred}(c) illustrates the results of cooperative QoS prediction for different size of the temporal segments. It shows that the NRMSE distance decreases considerably with the number of the temporal segments. When the number of segments is one, i.e., the similarity is computed based on the one-year workloads of the consumer, the prediction accuracy is very low. However, if the similarity is computed based on smaller segments, the prediction accuracy increases.

\subsubsection{Evaluation of QLIS approach}

We rank the providers based on the consumer's long-term QoS requirements and the predicted QoS performance of each provider. First, we rank the provider based on the actual QoS performance and expected QoS performance of the consumer. Next, we compare this rank with the proposed approaches, i.e., ranking based on cooperative prediction and long-term prediction without historical information. We consider two traditional ranking approaches based on 1) trial performance and 2) advertised performance \cite{wang2018testing}. Each rank is computed based on the NRMSE distances of the expected QoS performance and predicted QoS performance. Table \ref{tab:ranks} shows the ranks of the provider based on different methods. Each provider is identified numerically from 1 to 5. The actual ranking order is 5, 2, 3, 1, and 4. The cooperative ranking successfully predicts the rank of 3 providers (5, 2, and 3). The long-term prediction without history and the trial-based ranking approaches correctly rank two providers (5 and 3). The ranking of the providers based on the advertisements provides the most inaccurate ranking of the providers. The results of the rankings may change if we choose different months for the trial.

The confidence of the cooperative ranking is shown in Table \ref{tab:ranks} for each provider. The highest-ranked provider has the highest confidence as it has the least distance between the trial experience and the predicted performance. The first four providers in the cooperative rankings have high prediction confidence. The prediction for the provider 1 has the least confidence value. The reason is that the performance of the last provider varies to a large extent in the trial period. When the confidence threshold is set to below 1, the last two providers will be discarded from the cooperative rankings.

\begin{table}[ht]
\vspace{-4mm}
\caption{Ranking of IaaS Providers}
\begin{tabular}{|c|c|c|c|c|}
 \hline
 Actual & Cooperative & Long-term & Trial & Advertisements \\ [0.5ex] 
 Ranks & Ranks & Ranks & Ranks & Ranks \\ [0.5ex] 
  \hline
  \hline
 5 & 5 (0.46) & 5 & 5 & 1 \\
 2 & 2 (0.53) & 4 & 4 & 3 \\
 3 & 3 (0.58) & 3 & 3 & 5 \\
 1 & 4 (1.44) & 2 & 2 & 4 \\
 4 & 1 (2.03) & 1 & 1 & 2 \\
 \hline
 \end{tabular}
\label{tab:ranks}
\vspace{-5mm}
\end{table}

\subsection{Discussion}

Experiment results show that relying only on IaaS advertisements or the trial experience is inadequate for the long-term selection as it often leads the incorrect selection. Moreover, the results show that selecting appropriate trial workloads has a substantial impact on long-term performance discovery and selection. The results of the CLQP approach shows that long-term performance discovery could be improved considerably with the help of past trial users' experience. Based on these findings, we suggest that consumers should perform trials based on the long-term characteristics of their workloads instead of relying only on stress testing. The proposed framework could be utilized to generate trial workloads based on a consumer's workload for long-term performance discovery. The performance discovery will be improved considerably when many consumers would share their trial experience.

\section{Conclusion}
We propose a long-term IaaS provider selection framework to select the closest match IaaS provider according to a consumer's long-term requirements. The short-term trial periods offered by the IaaS providers are leveraged to discover the providers' unknown QoS performance. We devise a temporal skyline-based filtering method to limits the number of candidate IaaS providers for the trial periods. Experimental results show that the filtering method effectively reduces the number of candidate providers for the trial period. The proposed CLQP approach utilizes the experience of trial users to predict the performance of a provider for the consumer's long-term workloads with a confidence measure. Experimental results show that the CLQP approach can effectively measure the QoS performance. The proposed LQP-short approach discovers QoS performance without history using the proposed trial workload generation approach. Experimental results show that the LQP-short can effectively predict QoS performance with acceptable precision. Finally, a QoS-aware selection method is proposed to select the closest match provider where the provider's predicted performance closely matches the consumer's long-term requirements. Experimental results show that it ranks the providers successfully. The proposed framework may help many consumers to choose the right provider with limited performance information. The consumers do not need to wait for a long time to accumulate enough information to make an informed decision. In the future, we will extend this work for the dynamic workloads of consumers where an online-prediction approach is required.

% if have a single appendix:
%\appendix[Proof of the Zonklar Equations]
% or
%\appendix  % for no appendix heading
% do not use \section anymore after \appendix, only \section*
% is possibly needed

% use appendices with more than one appendix
% then use \section to start each appendix
% you must declare a \section before using any
% \subsection or using \label (\appendices by itself
% starts a section numbered zero.)
%

% \appendices
% \section{Proof of the First Zonklar Equation}
% Appendix one text goes here.

% % you can choose not to have a title for an appendix
% % if you want by leaving the argument blank
% \section{}
% Appendix two text goes here.

% % use section* for acknowledgment
% \ifCLASSOPTIONcompsoc
%   % The Computer Society usually uses the plural form
%   \section*{Acknowledgments}
% \else
%   % regular IEEE prefers the singular form
%   \section*{Acknowledgment}
% \fi

% The authors would like to thank...

% Can use something like this to put references on a page
% by themselves when using endfloat and the captionsoff option.
\ifCLASSOPTIONcaptionsoff
  \newpage
\fi

% trigger a \newpage just before the given reference
% number - used to balance the columns on the last page
% adjust value as needed - may need to be readjusted if
% the document is modified later
%\IEEEtriggeratref{8}
% The "triggered" command can be changed if desired:
%\IEEEtriggercmd{\enlargethispage{-5in}}

% references section

% can use a bibliography generated by BibTeX as a .bbl file
% BibTeX documentation can be easily obtained at:
% http://mirror.ctan.org/biblio/bibtex/contrib/doc/
% The IEEEtran BibTeX style support page is at:
% http://www.michaelshell.org/tex/ieeetran/bibtex/
%\bibliographystyle{IEEEtran}
% argument is your BibTeX string definitions and bibliography database(s)
%\bibliography{IEEEabrv,../bib/paper}

\bibliographystyle{IEEEtran}
% argument is your BibTeX string definitions and bibliography database(s)
\bibliography{IEEEabrv,Main}
%
% <OR> manually copy in the resultant .bbl file
% set second argument of \begin to the number of references
% (used to reserve space for the reference number labels box)
% \begin{thebibliography}{1}

% \bibitem{IEEEhowto:kopka}
% H.~Kopka and P.~W. Daly, \emph{A Guide to \LaTeX}, 3rd~ed.\hskip 1em plus
%   0.5em minus 0.4em\relax Harlow, England: Addison-Wesley, 1999.

% \end{thebibliography}

% biography section
% 
% If you have an EPS/PDF photo (graphicx package needed) extra braces are
% needed around the contents of the optional argument to biography to prevent
% the LaTeX parser from getting confused when it sees the complicated
% \includegraphics command within an optional argument. (You could create
% your own custom macro containing the \includegraphics command to make things
% simpler here.)
%\begin{IEEEbiography}[{\includegraphics[width=1in,height=1.25in,clip,keepaspectratio]{mshell}}]{Michael Shell}
% or if you just want to reserve a space for a photo:

\vspace{-12mm}
\begin{IEEEbiography}[{\includegraphics[width=1in,height=1.25in,clip,keepaspectratio]{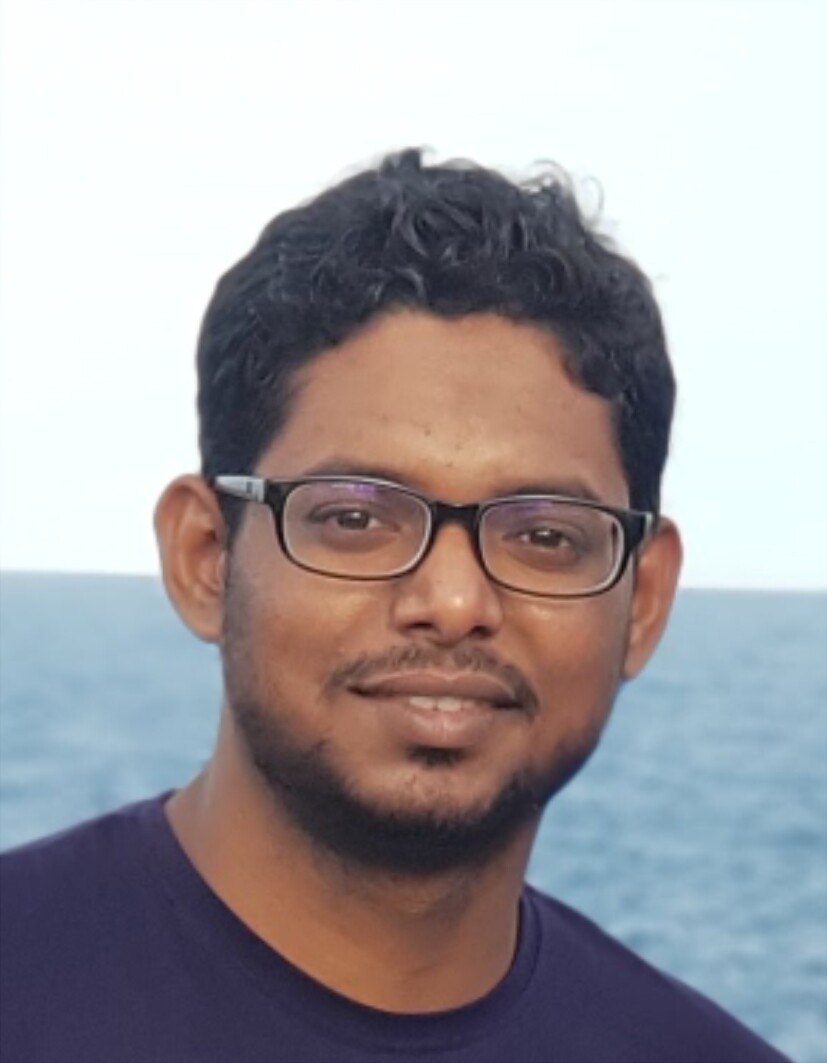}}]{Sheik Mohammad Mostakim Fattah} 
is a Ph.D. student in the School of Computer Science, University of Sydney, Australia. He obtained his Masters degree from Hankuk University of Foreign Studies, South Korea. He received his Bachelor degree in Computer Science and Engineering from University of Dhaka, Bangladesh. He was a research assistant in Korea Electronics Technology Institute. His research interests are Service Computing, Cloud Computing, Internet of Things, Web of Things, and Semantic Web.
\end{IEEEbiography}
\vspace{-12mm}
\begin{IEEEbiography}[{\includegraphics[width=1in,height=1.25in,clip,keepaspectratio]{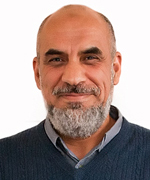}}]{Athman Bouguettaya}
is Professor and Head of School of Computer Science at University of Sydney, Australia. He received his PhD in Computer Science from the University of Colorado at Boulder (USA) in 1992. He was previously Science Leader in Service Computing at CSIRO ICT Centre, Canberra. Australia. Before that, he was a tenured faculty member and Program director in the Computer Science department at Virginia Polytechnic Institute and State University (commonly known as Virginia Tech) (USA). He is or has been on the editorial boards of several journals including, the IEEE Transactions on Services Computing, ACM Transactions on Internet Technology, the International Journal on Next Generation Computing and VLDB Journal. He has published more than 250 books, book chapters, and articles in journals and conferences in the area of databases and service computing (e.g., the IEEE TKDE, the ACM TWEB, WWW Journal, VLDB Journal, SIGMOD, ICDE, VLDB, and EDBT). He was the recipient of several federally competitive grants in Australia (e.g., ARC) and the US (e.g., NSF, NIH). He is a Fellow of the IEEE and a Distinguished Scientist of the ACM.
\end{IEEEbiography}
\vspace{-12mm}
\begin{IEEEbiography}[{\includegraphics[width=1in,height=1.25in,clip,keepaspectratio]{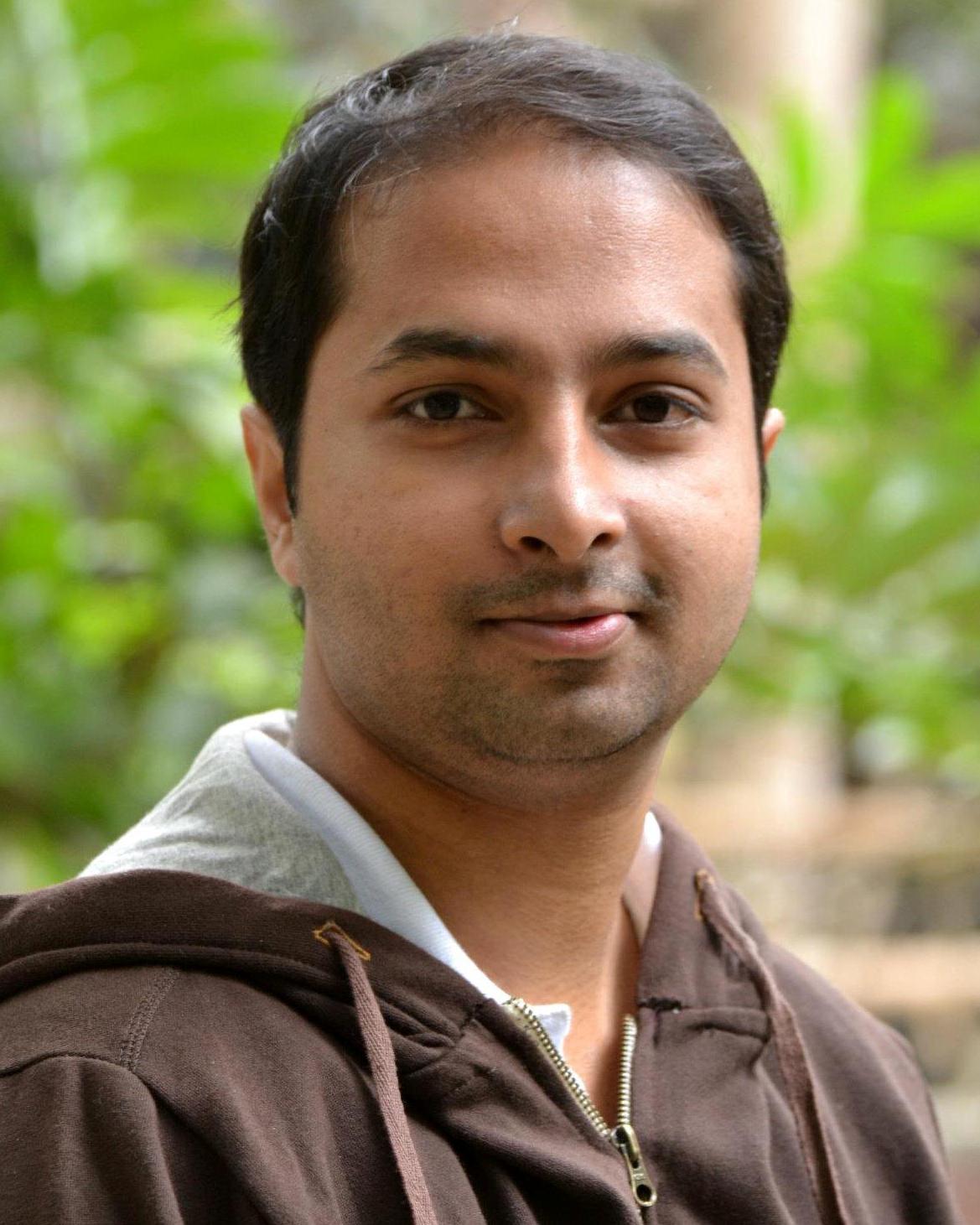}}]{Sajib Mistry} 
is Lecturer at School of Elect Eng, Computer and Math Sci in Curtin University, Australia. He was a postdoc fellow at School of Computer Science in University of Sydney. He received his PhD from the RMIT University, Australia. His research interests include Edge/Cloud Computing, Big Data and IoT. He has published articles in international journals and conferences, such as IEEE TSC, TKDE, ACM CACM, ICSOC, WISE, and ICWS. He received the best paper award in ICSOC2016.
\end{IEEEbiography}

% \begin{IEEEbiography}{Michael Shell}
% Biography text here.
% \end{IEEEbiography}

% % if you will not have a photo at all:
% \begin{IEEEbiographynophoto}{John Doe}
% Biography text here.
% \end{IEEEbiographynophoto}

% % insert where needed to balance the two columns on the last page with
% % biographies
% %\newpage

% \begin{IEEEbiographynophoto}{Jane Doe}
% Biography text here.
% \end{IEEEbiographynophoto}

% You can push biographies down or up by placing
% a \vfill before or after them. The appropriate
% use of \vfill depends on what kind of text is
% on the last page and whether or not the columns
% are being equalized.

%\vfill

% Can be used to pull up biographies so that the bottom of the last one
% is flush with the other column.
%\enlargethispage{-5in}

% that's all folks
\end{document}